\documentclass[12pt]{article}
\usepackage{amsfonts,amssymb,amsbsy,amsmath,amsthm,mathrsfs,enumerate,verbatim}
\usepackage[mathscr]{eucal}

 \usepackage[colorlinks=true]{hyperref}
 \usepackage{graphicx}

 \hypersetup{urlcolor=blue, citecolor=red}
\newtheorem{theo}{Theorem}[section]
\newtheorem{prop}[theo]{Proposition}

\newtheorem{defi}[theo]{Definition}
\theoremstyle{definition}
\newtheorem{rema}[theo]{Remark}

\newcommand{\bel}{\begin{equation} \label}
\newcommand{\ee}{\end{equation}}

\newcommand{\eps}{{\varepsilon}}
\def\beq{\begin{equation}}
\def\eeq{\end{equation}}
\newcommand{\bea}{\begin{eqnarray}}
\newcommand{\eea}{\end{eqnarray}}
\newcommand{\beas}{\begin{eqnarray*}}
\newcommand{\eeas}{\end{eqnarray*}}

{

\begin{document}

\begin{center}
{\Large \bf Spectral monodromy of small non-selfadjoint perturbed operators: completely integrable
or quasi-integrable case }

\medskip
(\today, provisional version)

\end{center}

\medskip

\begin{center}
{\footnote{Institute of Mathematics, Faculty of Mathematics and Computer Science, Jagiellonian
University, 30-348 Krak\'ow, Poland. //
E-mail: quang.phan@uj.edu.pl}{Quang Sang PHAN}}
\end{center}

\begin{abstract} ~\\
The spectral monodromy is a combinatorial invariant, defined directly from the spectrum of certain non-selfadjoint classical operators. It is a obstruction against the global lattice structure of the spectrum, seen as a discrete subset of points in the complex plane, and in the semi-classical limit.
We work with small non-selfadjoint perturbations of classical selfadjoint operators with two degrees of freedom, assuming that the (semi-)classical principal symbol of the unperturbed part is in two different cases: completely integrable system in the first case, and quasi-integrable one with a globally (non-degenerate) isoenergetic condition in the second case.
In each case, the corresponding spectral monodromy allows to recover the corresponding classical monodromy.

\end{abstract}

\medskip

% {\bf  AMS 2010 Mathematics Subject Classification:} 35R30.\\

~

\tableofcontents

%%%%%%%%%%%%%%%%%%%%%%%%%%%%%%%%%%%%%%%%%%%%%%%%%%%%%%%%%%%%%%%%%%%%%%%%%%%%%%%%%%%%%%%%%%%%%%%%%%%%%%%%%%%%%%%%%%%%%%%%%%%%%%%%%%%%%%%%%%%%%%%%%%%%%%%%%%%%%%%%%%%%%%%%%%%%%%%%%%%
%%%%%%%%%%%%%%%%%%%%%%%%%%%%%%%%%%%%%%%%%%%%%%%%%%%%%%%%%%%%%%%%%%%%%%%%%%%%%%%%%%%%%%%%%%%%%%%%%%%%%%%%%%%%%%%%%%%%%%%%%%%%%%%%%%%%%%%%%%%%%%%%%%%%%%%%%%%%%%%%%%%%%%%%%%%%%%%%%%%
%%%%%%%%%%%%%%%%%%%%%%%%%%%%%%%%%%%%%%%%%%%%%%%%%%%%%%%%%%%%%%%%%%%%%%%%%%%%%%%%%%%%%%%%%%%%%%%%%%%%%%%%%%%%%%%%%%%%%%%%%%%%%%%%%%%%%%%%%%%%%%%%%%%%%%%%%%%%%%%%%%%%%%%%%%%%%%%%%%%

\section{Introduction and motivation}

% définir une propriété de monodromie spectrale associée au spectre, vu comme un ensemble discret de points de C.

%   a property of monodromy associated spectral spectrum, seen as a discrete set of points of C.

We are interested in the global structure of the spectrum of non-selfadjoint $h-$Weyl-pseudodifferential operators with two degrees of freedom, in the semi-classical limit, that is, when the small classical parameter $h$ tends to $0$. We will define in this paper a feature-the \emph{spectral monodromy} associated to the spectrum, seen as a discrete subset of points in the complex plane.
The considered operators are small non-selfadjoint perturbations of selfadjoint operators, with certain different assumptions on the classical dynamic of the unperturbed part.

The spectral monodromy was defined the first time in \cite{QS14} for classes of classical operators, associated with completely integrable systems.
In that work, one assumed that, the principal symbol of the unperturbed part and the leading term of the perturbed part are in involution for the Poisson bracket. This is a particular case of the present work which is more interesting and quite developed. We assume in this paper that the classical flow of the unperturbed part is in two cases: completely integrable in the first case, and quasi-integrable with a globally non-degenerate condition in the second case.

In the paper \cite{Broer07}, the authors succeeded to define a monodromy related to quasi-integrable systems. They proposed a mysterious question: whether a quantum invariant (like monodromy) can be defined for classical operators in the quasi-integrable case. This is also a motivation of our work. The results obtained in the second case of this paper answer completely that question.

We give now a brief description for the monodromy spectral.
The theory of spectral asymptotics (for example Refs. \cite{Hitrik04}, \cite{Hitrik05}, \cite{Hitrik08}, and especially \cite{Hitrik07}) showed that under certain technical general hypothesis, the spectrum of such operators in domains depending on the classical parameter of the complex
spectral plane has microlocally a deformed lattice structure. That means it is image of a square lattice by a sort of local chart, called a \textit{micro chart}. Such lattices are of small size but they can be defined around many points in the spectral domain. A family of close micro-lattices in a small domain forms a local lattice, on which we have a corresponding family of micro-charts, called a \textit{pseudo-chart}. The pseudo-charts play the role of local charts on the spectral domain.

A natural question we propose is that: whether the spectrum has globally a lattice structure? The spectral monodromy a combinatorial invariant, related to the spectrum, allows us to answer this question. It is given when we analyze how the local spectral lattices are glued, by examining
transition maps between overlapping pseudo-charts. In fact, these local lattices are glued them-self by a special structure, which is characterized by the spectral monodromy. It is defined as a element of the \v{C}ech cohomology group $\check{H}^1(U,GL(2, \mathbb Z) )$, independent of classical parameter and given small perturbations. When the spectral monodromy is not trivial, then the transition maps either, and the spectrum hasn't a smooth global lattice structure.

There exists a known quantum invariant, that is quantum monodromy given by Vu Ngoc, see \cite{Vu-Ngoc99}. It is defined for the joint spectrum of integrable quantum systems of $ n \geq 2$ commuting selfadjoint classical operators. Contrariwise, the spectral monodromy is defined for a single non-selfadjoint operator. However the quantum monodromy stills useful for our work. It give simple examples of the existence of monodromy for the spectrum. In fact, when a non-selfadjoint operators is normal, then its discrete spectrum is completely identified with the joint spectrum of the integrable quantum system, which is composed of the real part and the imaginary part of the operator. For more details on this point, we refer to \cite{QS14} Sec. II.
The existence of non-trivial monodromy is known and assured for several integrable quantum systems, for example the quantum spherical pendulum \cite{Cush88}, \cite{Guil89} or the quantum Champagne bottle \cite{Child98}, \cite{Bates91}.

In both cases, we consider Weyl-$h-$pseudodifferential non-selfadjoint perturbed operator in dimension $2$, depending on a small parameter $\varepsilon$ such that $h \ll \varepsilon = \mathcal{O}(h^\delta)$, with $0< \delta <1 $.

In the first case, we treat such operators of the form $ P_{\varepsilon}= P(x,hD_x,\varepsilon; h ),$
where the unperturbed operator $P:= P_{\varepsilon=0}$ is formally selfadjoint, assuming that the principal classical symbol $p$ of $P$ is completely integrable. A particular case of this case when the perturbed operators of the form  $P_\varepsilon= P+ i
\varepsilon Q$, assuming that the corresponding principal symbols $p$ of $P$ and $q $ of $Q$ commute for the Poisson bracket, is given in \cite{QS14} Sec. III.

In the second case, we perturbs even the completely integrable symbol $p$ by a small quantity to get a quasi-integrable system.
We consider classical operators $P_{\varepsilon, \lambda}$, depending also smoothly on a small enough parameter $0< \lambda \ll 1$, and assuming that the principal symbol of the selfadjoint unperturbed operators $P_{\lambda}:=P_{\varepsilon =0, \lambda}$ is a quasi-integrable system of the form $p_\lambda = p+ \lambda p_1$, here $p_1$ is a bounded Hamiltonian. Moreover, a general hypothesis given is globally isoenergetic non-degeneracy of $p$.

It knows from the spectral asymptotic theory \cite{Hitrik07} that, under an ellipticity condition at infinity (see \eqref{el con}), in both cases, the spectrum of perturbed operators is discrete, and included in a horizontal band of size $\mathcal{O}(\varepsilon)$. Moreover, that work allows give an asymptotic expansion of the eigenvalues, located in a some small domains of size $\mathcal{O} (h ^\delta) \times \mathcal{O }(\varepsilon h^\delta)$ of the spectral band, called good rectangles. These good rectangles are associated with
Diophantine invariant Lagrangian tori in the phase space, on which the Hamiltonian flow of the unperturbed part, either $p$ in the first case or $ p_\lambda $ in the second case, is quasi-periodic of constant frequencies.

On the other hand, the existence of such tori is insured and moreover the tori form a set of Cantor type of full measure (in measure theoretic sense) in the phase space. In the first case, it is classical since the fact that the phase space of the completely integrable Hamiltonian system $p$ is foliated by invariant Lagrangian $2$-tori (the Liouville tori) and almost of them satisfy a Diophantine condition. And then, for the seconde case, we need the KAM theory which deals quasi-periodic motions in small perturbed Hamiltonian systems (like $p_\lambda $). It shows that, under non-degeneracy assumptions and $\lambda$ small enough, these Diophantine invariant tori for the flow of $p$ persist (slightly deformed) as Diophantine invariant tori (KAM tori) for one of the perturbed system $ p_\lambda $.

Therefore, we have a lot of good rectangles in the spectral band. The corresponding spectrum $\mu$ of the perturbed operators in each good rectangle, denoted by $R^{(a)}(\varepsilon,h)$, here $a \in \mathbb R^2$ is used to fix the good rectangle, is given by
\begin{eqnarray} \label{s}
                  R^{(a)}(\varepsilon,h) \ni \mu
                        & = & P \Big (\xi_a+ h(k-\frac{k_1}{4} )-\frac{S_1}{2 \pi}, \varepsilon;h \Big )
                        + \mathcal O(h^\infty), \quad k \in \mathbb Z^2,
\end{eqnarray}
uniformly for small $h, \varepsilon $. Here $\xi_a$ are action coordinates, $S_1 \in \mathbb R^2$ is the
action, $k_1 \in \mathbb Z^2$ is the Maslov index of the fundamental cycles, of the corresponding Diophantine torus. $P(\xi,\varepsilon; h )$ is a smooth function admitting asymptotic expansion in
$( \xi,\varepsilon, h )$.

The spectrum so has the structure of a deformed lattice, with horizontal spacing $h$ and vertical spacing $\varepsilon h$.
Since \eqref{s}, there is a local diffeomorphism, denoted by $f$, that sends the spectrum in the rectangle to a part of
 $h \mathbb Z^2 $, with modulo $\mathcal O(h^\infty)$,
 \begin{eqnarray}
    R^{(a)}(\varepsilon,h) \ni  \mu  \mapsto    f(\mu;\varepsilon, h) \in h \mathbb Z^2 +\mathcal O(h^\infty).
\end{eqnarray}
This map is called a micro-chart. A family (of Cantor type) of close micro-charts on a small domain forms a local pseudo-chart of the spectrum, as discussed at the beginning of this section.

We notice that each micro-chart is normally valid for one good rectangle. However, an important fact we show in this paper that we can build spectral pseudo-charts such that the leading term in asymptotic expansions of their micro-charts in small parameters $h, \varepsilon $ is locally well defined.
This leading term is well defined for every micro-chart of a pseudo-chart.
The construction of micro charts in each case will respectively be done in detail in Sec. \ref{sp-lattice Sec} and Sec. \ref{end mono}.

With regard to the global problem, we consider the spectrum as a discrete subset of the complex plane. We can apply a result of \cite{QS14} about a discrete set, called asymptotic pseudo-lattice (see also Definition \ref{pseu-lattice}) for the spectrum. It shows that the differential of the transition maps between two overlapping local pseudo-charts $ (f_i, U^i(\varepsilon)) $ and $ (f_ j, U^j (\varepsilon))$ is in the group $GL(2,\mathbb Z)$, with modulo $\mathcal O(\varepsilon, \frac{h}{\varepsilon})$:
    $$ d (\widetilde{f}_i) = M_{ij} d( \widetilde{f}_j)+ \mathcal O(\varepsilon, \frac{h}{\varepsilon}),$$
with $\widetilde{f}_i= f_i \circ \chi$, $\widetilde{f}_i= f_i \circ \chi$, where $\chi$ is the function $ (u_1,u_2) \mapsto (u_1,\varepsilon u_2)$, and $M_{ij} \in GL(2, \mathbb Z)$ is a integer constant matrix.

Let $U (\varepsilon)$ be a bounded open domain in the spectral band and cover it by an arbitrary (small enough) locally finite covering of pseudo-charts
$ \{ \left(  f_ j ,  U^j (\varepsilon) \right) \}_{j \in \mathcal{J} }$, here $\mathcal{J}$ is a finite index set. Then the spectral monodromy on $U (\varepsilon)$ is defined as the unique $1$-cocycle $\{  M_{ij} \} $, with modulo-coboundary in the first \v{C}ech cohomology group.
We have just explained the principe idea how to define the monodromy.
~\\

This work is a inverse quantum problem. The spectral monodromy is defined directly from the spectrum. However, as an another important result, it is strictly related to the classical problem, and the classical results, in turn, illuminate again the initial quantum problem.
In both cases, the spectral monodromy allows us to recover known classical invariants. In the first case, the monodromy spectral can be identified to the classical monodromy for the Liouville invariant tori of integrable systems, given in \cite{Duis80}. And in the second case, the spectral monodromy allows to recover the monodromy of the KAM invariant tori of quasi-integrable systems, defined by Broer and co-workers \cite{Broer07}. Then the geometry of the corresponding principal symbols plays an important role in quantum properties of classical operators.

%%%%%%%%%%%%%%%%%%%%%%%%%%%%%%%%%%%%%%%%%%%%%%%%%%%%%%%%%%%%%%%%%%%%%%%%%%%%%%%%%%%%%%%%%%%%%%%%%%%%%%%%%%%%%%%%%%%%%%%%%%%%
%%%%%%%%%%%%%%%%%%%%%%%%%%%%%%%%%%%%%%%%%%%%%%%%%%%%%%%%%%%%%%%%%%%%%%%%%%%%%%%%%%%%%%%%%%%%%%%%%%%%%%%%%%%%%%%%%%%%%%%%%%%%%%%

\section{Spectral monodromy in the completely integrable case}
% \section{Pseudodifferential operators}

In this first case, we shall define the spectral monodromy, associated with the spectrum of small non-selfadjoint perturbations of selfadjoint classical operators in two dimensions, assuming that the classical flow of the unperturbed part is completely integrable.

We will work throughout this article with pseudodifferential operators obtained by the $h-$Weyl-quantization of a standard space of symbols on $T^*M =\mathbb R^{2n}_{(x,\xi)}$, here $M= \mathbb R^n$ or a manifold compact of $n$ dimensions, and in particulary $n=2$. We denote $\sigma $ the standard $2-$ symplectic form on $T^*M$.

In the following, we introduce classical operators on $M= \mathbb R^n$, but it is alright general to the manifold case.

        \begin{defi} \label{fonc ord}
                        A function $m: \mathbb R^{2n} \rightarrow (0, + \infty)$ is called an order function
                     if there are constants  $C,N >0$ such that
                                $$m(X)  \leq C \langle X-Y\rangle^{ N} m(Y), \forall X,Y \in \mathbb R^{2n},$$
        with notation $\langle Z\rangle= (1+ |Z|^2)^{1/2}$ for $Z \in \mathbb R^{2n}$.
        \end{defi}

% One use often the order function  $m(Z) \equiv 1$ or $$m(Z)= \langle Z \rangle ^{l/2}= (1 + |Z|^2 )^{l/2},$$ with a given constant $l \in \mathbb R $.

         \begin{defi}
                        Let $m$ be an order function and $k \in \mathbb R$, we define classes of symbols of $h$-order $k$, $S^k(m)$ (families of functions) of $(a(\cdot;h))_{h \in (0,1]}$ on $\mathbb R^{2n}_{(x,\xi)}$ by
                        \begin{equation}
                                S^k(m)= \{ a \in C^\infty (\mathbb R^{2n})
                                 \mid  \forall \alpha \in \mathbb N ^{2n}, \quad |\partial^\alpha a | \leq  C_\alpha h^k m \} ,
                        \end{equation}
         for some constant $C _\alpha >0$, uniformly in $h \in (0,1]$. \\
         A symbol $a$ is called $\mathcal O(h^\infty)$ if it's in $\cap _{k \in \mathbb R } S^k(m):= S^{\infty}(m) $.
        \end{defi}

        Then $ \Psi^k(m)(M)$ denotes the set of all (in general unbounded) linear operators $A_h$ on $L^2(\mathbb R^n)$, obtained from the $h-$Weyl-quantization of symbols $a(\cdot;h) \in S^k(m) $ by the integral:
        \begin{equation} \label{symbole de W}
                            (A_h u)(x)=(Op^w_h (a) u)(x)= \frac{1}{(2 \pi h)^n}
                                 \int_{ \mathbb R^{2n}} e^{\frac{i}{h}(x-y)\xi}
                                 a(\frac{x+y}{2},\xi;h) u(y) dy d\xi.
        \end{equation}

We refer to Refs. \cite{Dimas99}, \cite{Robert87}, and \cite{Shubin01} for the theory of classical operators.
In this paper, we always assume that symbols admit a classical asymptotic expansion in integer powers of $h$.
The leading term in this expansion is called the principal symbol of operators.

%%%%%%%%%%%%%%%%%%%%%%%%%%%%%%%%%%%%%%%%%%%%%%%%%%%%%%%%%%%%%%%%%%

\subsection{Spectral asymptotic for small perturbed non-selfadjoint operators}  \label{s a}
In this section, we recall some results of the spectral asymptotic theory for small non-selfadjoint perturbations of selfadjoint classical operators in two dimensions.

    \subsubsection{General assumptions}  \label{hypothese}

 We give here general assumptions of small non-selfadjoint perturbations of selfadjoint operators in two dimensions, as in Refs. \cite{Hitrik07}, \cite{Hitrik04}, and \cite{Hitrik05}.

    $ M $ denotes $\mathbb R^2$ or a connected compact analytic real (Riemannian) manifold of dimension $ 2 $ and we denote by $\widetilde{M}$
    the canonical complexification of $ M $, which is either $ \mathbb C ^ 2 $ in the Euclidean case or a Grauert tube in the case of manifold
    (see Ref. \cite{Burns01}).

%%%%%%%%%%%%%%%%%%%%%%%%%%%%%%%%%%    REFERENCE CHO Grauert tube??
%%%%%%%%%%%%%%%%%%%%%%%%%%%%%%%%%%

    We consider a non-selfadjoint $ h -$ pseudodifferential operator $P_{\varepsilon}$ on $ M $ and suppose that
    \begin{equation}
         P_{\varepsilon=0}:= P \quad \textrm{is formally self-adjoint}.
    \end{equation}

    Note that if $ M = \mathbb R ^ 2 $, the volume form $ \mu (dx) $ is naturally induced by the Lebesgue measure on $ \mathbb R ^ 2 $.
    If $ M $ is a compact Riemannian manifold, then the volume form $ \mu(dx) $ is induced by the given Riemannian structure of $ M $.
    Therefore in both cases the volume form is well defined and the operator $P_{\varepsilon}$ may be seen as an (unbounded) operator
    on $L^2(M, \mu(dx)) $.
    We always denote the principal symbol of $ P_ {\varepsilon} $ by $ p_ \varepsilon $ which is defined on $ T ^ * M $.

    We will assume the ellipticity condition at infinity for $ P_ {\varepsilon} $ at some energy level $E \in \mathbb R$ as follows:

    When $M=\mathbb R^2$, let
                        \begin{equation}
                            P_{\varepsilon}= P(x,hD_x,\varepsilon; h )
                        \end{equation}
   be the Weyl quantification of a total symbol $P(x, \xi,\varepsilon; h )$ depending smoothly on $\varepsilon$ in a neighborhood of $(0, \mathbb R) $ and taking values in the space of holomorphic functions of $(x,\xi)$ in a tubular neighborhood of $\mathbb R^4$ in $\mathbb C^4$ on which we assume that:
                \begin{equation}
                           | P(x, \xi,\varepsilon; h ) | \leq  \mathcal O(1) m(Re(x,\xi)).
                \end{equation}
    Here $ m $ is an order function in the sense of Definition \ref{fonc ord}.
    We assume moreover that $ m> 1 $ and $ P_ {\varepsilon} $ is classical of order $0$,
     \begin{equation}
        P(x, \xi,\varepsilon; h ) \sim \sum_{j=0}^\infty
                             p_{j,\varepsilon}(x,\xi) h^j, h \rightarrow 0,
     \end{equation}
     in the selected space of symbols. \\
     In this case, the main symbol is the first term of the above expansion, $p_\varepsilon = p_{0,\varepsilon}$
     and the ellipticity condition at infinity is
                        \begin{equation}    \label{el con}
                            |p_{0,\varepsilon}(x,\xi) - E| \geq \frac{1}{C} m(Re(x,\xi)), \mid (x,\xi)\mid \geq C,
                        \end{equation}
     for some $ C> 0 $ large enough.

     When $ M $ is a compact manifold, we consider $ P_ \varepsilon $ a differential operator on $ M $ such that in local coordinates $ x $ of $ M $, it is of the form:
        \begin{equation}
                            P_\varepsilon = \sum_{|\alpha |\leq m} a_{\alpha,\varepsilon}(x;h)(hD_x)^\alpha,
        \end{equation}
  where $D_x= \frac{1}{i} \frac{\partial}{\partial x}$ and $a_{\alpha,\varepsilon}$ are smooth functions of $\varepsilon$ in a neighborhood of $0$
   with values in the space of holomorphic functions on a complex neighborhood of $x=0$. \\
  We assume that these $a_{\alpha,\varepsilon}$ are classical of order $0$,
                         \begin{equation}
                           a_{\alpha,\varepsilon}(x;h) \sim \sum_{j=0}^\infty
                             a_{\alpha,\varepsilon,j}(x) h^j, h \rightarrow 0,
                        \end{equation}
  in the selected space of symbols. \\
  In this case, the principal symbol $p_\varepsilon$ in the local canonical coordinates associated $(x,\xi) $ on $T^*M $ is \begin{equation}
                           p_\varepsilon(x,\xi)= \sum_{|\alpha | \leq m} a_{\alpha,\varepsilon,0}(x) \xi^{\alpha}
  \end{equation}
  and the elipticity condition at infinity is
            \begin{equation}
                            |p_{\varepsilon}(x,\xi) -E | \geq \frac{1}{C} \langle \xi \rangle ^m, (x,\xi) \in T^*M, \mid \xi \mid \geq C,
            \end{equation}
  for some $C>0$ large enough.
  Note here that $ M $ has a Riemannian metric, then $\mid \xi \mid$ and $\langle \xi \rangle= (1+  \mid \xi \mid ^2)^{1/2}$ are defined.

  It is known from Refs. \cite{Hitrik07}, and \cite{Hitrik04} that with the above conditions, the spectrum of $ P_ \varepsilon $ in
  a small but fixed neighborhood of $ E$ in $\mathbb C $ is discrete,
  when $h>0, \varepsilon \geq 0$ are small enough. Moreover, this spectrum is contained in a horizontal band of size $\varepsilon$:
    \begin{equation} \label{band}   |\mathrm{Im} (z)| \leq \mathcal O(\varepsilon ).\end{equation}

Let $p= p_{\varepsilon=0}$, it is principal symbol of the selfadjoint unperturbed operator $ P $
and therefore real. And let $q=\frac{1}{i}(\frac{\partial}{\partial \varepsilon})_{\varepsilon
                    =0}p_\varepsilon$ and assume that $q$ is a bounded analytic function on $T^*M$.
We can write the principal symbol
                            \begin{equation}  \label{symb prin}
                                  p_\varepsilon=p+i \varepsilon q+ \mathcal O (\varepsilon ^2).
                            \end{equation}

% In this case, we  the assumptions on the classical flow of unperturbed principal symbol.

We assume that $ p $ is completely integrable, i.e., there exists an analytic real valued function
$f$, differentially independent of $p$ such that $ \{p, f \}= 0$ with respect to the Poisson
bracket on $T^*M$. That means
\begin{equation} \label{F}
F=(p,f): T^*M \rightarrow \mathbb R^2 \end{equation}
is a momentum map.
Then the space of regular leaves of $F$ is foliated by Liouville Lagrangian invariant tori by
the angle-action Theorem \ref{A-A}.

We assume also that
  \begin{equation}p^{-1}(E) \cap T^*M  \ \textrm{is connected}, \end{equation}
and the energy level $ E $ is regular for $ p $, i.e., $dp \neq 0$ on $p^{-1}(E) \cap T^*M$.
The energy space $p^{-1}(E)$ is decomposed into
a singular foliation:
 \begin{equation} p^{-1}(E) \cap T^*M   = \bigcup_{a \in J} \Lambda_a ,\end{equation}
 where $ J $ is assumed to be a compact interval, or, more generally, a connected graph with a finite number of vertices and of edges, see pp. 21-22 and 55 of Ref. \cite{Hitrik07}.

We denote by $S$ the set of vertices.
For each $a \in J$, $\Lambda_a$ is a connected compact subset invariant with respect to $H_p$.
Moreover, if $a \in J\backslash S$, $\Lambda_a$ is a invariant Lagrangian torus depending
analytically on $a$. These tori are regular leaves corresponding regular values of $F$.
Each edge of $ J $ can be identified with a bounded interval of $ \mathbb R $ and we have therefore
a distance on $J$ in the natural way.

% Next, we assume the continuity of tori: let $\Lambda_{a_0}, a_0 \in  J\backslash S$, for all $\mu >0, \exists \ \gamma >0$, such that if $dist(a,a_0) %  < \gamma $, then $\Lambda_a \subset \{\rho \in p^{-1}(E) \cap T^*M: dist(\rho, \Lambda_0) < \mu \}$.

% Note that this hypothesis holds for integrable systems with non-degenerate singularities.
% day la hypothesis ma chung ta se can den trong truong hop quasi-integrable.

We denote $H_p$ the Hamiltonian vector field of $p$, defined by $\sigma (H_p, \cdot)= -dp (\cdot)$.
For each $a \in J$, we define a compact interval in $\mathbb R$:
                            \begin{equation} \label{Q vo cung}
                                    Q_\infty(a)=
                                     \big [ \lim_{T\rightarrow \infty} \inf_{\Lambda_a} Re \langle q \rangle _T,
                                      \lim_{T\rightarrow \infty} \sup_{\Lambda_a} Re \langle q \rangle _T\big],
                            \end{equation}
where $\langle q \rangle _T$, for $T>0$, is the symmetric average time $T$ of $q$ along the
$H_p-$flow, defined by
                            \begin{equation}  \label{t-average}
                                 \langle q \rangle _T= \frac{1}{T} \int_{-T/2}^{T/2} q \circ exp(t H_p) dt.
                            \end{equation}
Then it is more detail than \eqref{band}, the spectrum the of $P_\varepsilon$ in the neighborhood of $E $ in $\mathbb C $ is located in the band
                      \begin{equation}  \label{loca. spectre 2}
                           \mathrm{Im} \left( \sigma(P_\varepsilon) \cap \{z \in \mathbb C: |Re z -E| \leq \delta \} \right) \subseteq
                             \varepsilon \big [ \inf \bigcup_{a \in J}Q_\infty(a)-o(1),
                              \sup \bigcup_{a \in J}Q_\infty(a) +  o(1) \big ],
                      \end{equation}
when $\varepsilon, h, \delta \rightarrow 0$ (see Ref. \cite{Hitrik07}).

Each invariant Lagrangian torus $\Lambda_a$, with $ a  \in J\backslash S$, locally can be embedded
in a Lagrangian foliation of $H_p-$invariant tori. By the angle-action Theorem \ref{A-A}, there are
analytic local angle-action coordinates $\kappa= (x, \xi) $ on an open neighborhood $V$ of $\Lambda_a$ in
$T^*M$,
\begin{equation}  \label{coor} \kappa= (x,\xi): V \rightarrow \mathbb T ^2 \times  A, \end{equation}
with $A$ is an open neighborhood of some $\xi_a \in \mathbb R^2$, such that $\Lambda_a $ is symplectically
identified with $\mathbb T^2 \times \{ \xi_a \}$ by $\kappa$, and $p$ becomes a function of action variables
$\xi$,
           \begin{equation} \label{p}
                    p\circ \kappa^{-1} =p(\xi)= p(\xi_1, \xi_2), \ \xi \in A.           \end{equation}

Let $\Lambda \subset V $ be an arbitrary invariant Lagrangian torus (close to $\Lambda_a$) and suppose that by $\kappa$, it is symplectically sent to the torus $\mathbb T^2 \times \{ \xi\} $, denoted by $\Lambda_{\xi}$, with some $\xi \in A$.
We introduce the following notation used throughout this paper.
\begin{equation}  \label{Lam}
\Lambda \simeq \Lambda_{\xi}.
\end{equation}
Then the frequency of the torus $\Lambda$ (also of $\Lambda_{\xi}$) is defined by
\begin{equation} \label{frequence}
\omega(\xi)= \frac{\partial p}{\partial \xi} (\xi)= \big ( \frac{\partial p}{\partial \xi_1} (\xi),
 \frac{\partial p}{\partial \xi_2} (\xi) \big ), \ \xi \in A .
 \end{equation}
% Sometimes $\omega(\xi)$ is seen as an element of $\mathbb R$. \\
In particular, the frequency $\Lambda_a$ is $\omega(\xi_a)= \frac{\partial p}{\partial
\xi} (\xi_a)$. It knows that $\omega$ depends analytically on $\xi \in A$. In particular, the
restriction $\omega(\xi_a)$ depends analytically on $\xi_a$, when $a \in J\backslash S$. We will
assume that the function $a \mapsto \omega(\xi_a)$ is not identically constant on any connected
component of $J \backslash S$.

From now, for simplicity, we will assume that $q$ is \textbf{real valued} (in the general case, simply replace $q$ by its real part $\mathrm{Re}(q)$).

We define the average of $ q $ on the torus $\Lambda$, with respect to the natural Liouville measure on $\Lambda$, denoted by $\langle q \rangle_{\Lambda_ \xi} $, as following
                      \begin{equation} \label{moyenne de q}
                       \langle q \rangle_{\Lambda}= \int_{\Lambda}q .\end{equation}

\begin{rema}
 In the action-angle coordinates $(x,\xi)$ given by \eqref{coor}, we have
 \begin{equation} \label{moyenne2}
 \langle q \rangle_{\Lambda} =  \langle q \rangle_{\Lambda_ \xi}= \langle q \rangle (\xi)=
 \frac{1}{(2\pi)^2}\int_{\mathbb{T}^2}q(x,\xi)dx, \ \xi \in A.
 \end{equation}
In particular, $\langle q \rangle_{\Lambda_a}=\langle q \rangle(\xi_a)$.
\end{rema}

\begin{rema}[see pp. 56-57 of Ref. \cite{Hitrik07}]  \label{rem1}
For $a \in J\backslash S$, if $\omega(a) \notin \mathbb{Q}$, that means the frequency $\omega(a)$
is non resonant, then along the torus $\Lambda_a$, the Hamiltonian flow of $p$ is ergodic. Hence the
limit of $\langle q \rangle _T$, when $ T \rightarrow \infty$ exists, and is
equals to the space average of $q$ over the torus, $ \langle q \rangle_{\Lambda_a}$. Therefore we
have $$Q_\infty(a)= \{\langle q \rangle_{\Lambda_a}\}.$$
\end{rema}

It is true that $\langle q \rangle_{\Lambda_a} $ depends analytically on $a \in J\backslash S$ and
we assume it can be extended continuously on $J$. Furthermore, we assume that the function $a
\mapsto \langle q \rangle_{\Lambda_a} = \langle q \rangle(\xi_a)$ is not identically constant on
any connected component of $J\backslash S$.

Assume furthermore that the differential of the functions $p(\xi)$ given in \eqref{p}, and of
$\langle q \rangle$ given in \eqref{moyenne2} are $\mathbb R-$linearly independent when
$\xi=\xi_a$.
Then $\langle q \rangle $ and $p$ are in involution in the neighborhood $V$ of $\Lambda_a$, due to
$\langle q \rangle $ is invariant under the flow of $p$.

% Mo rong $\langle q \rangle $ thanh 1 ham smooth tren toan bo ko gian??

%%  Chung ta di nhien co the xem ham nay la ham cua a \in J/S, hoac \xi trong lan can cua \xi_a hoac tham tri tren ko gian phase o lan
%%%% can cua torus Lambda_a.

%%%%%%%%%%%%%%%%%%%%%%%%%%%%%%%%%%%%%%%%%%%%%%%%%%%%%%%%%%%%%%%%%%%%%%%%%%%%%%%
%%%%%%%%%%%%%%%%%%%%%%%%%%%%%%%%%%%%%%%%%%%%%%%%%%%%%%%%%%%%%%%%%%%%%%%%%%%%%%%%

\subsubsection{Asymptotic expansion of eigenvalues}

The asymptotic spectral theory (see Refs \cite{Hitrik04}- \cite{Hitrik07}) allows us to give an asymptotic description of all the eigenvalues
of $P_\varepsilon$ in some adapted small complex windows of the spectral band, which are associated with
Diophantine tori in the phase space. The force of the perturbation $\varepsilon $ is small and can be dependent or independent of the classical parameter $h$.
In our work we present the result in the case when $\varepsilon$ is sufficiently small, dependent
on $h$, and in the following regime
        $$h \ll \varepsilon = \mathcal{O}(h^\delta), $$
 where $\delta >0$ is some number small enough but fixed.
In this case, the spectral results are related to $(h,\varepsilon)$-dependent small windows.
\begin{defi}  \label{diop}
 Let $\alpha >0 $, $d>0$, and $\Lambda \simeq \Lambda_{\xi} $ be a $H_p-$ invariant Lagrangian torus, as in \eqref{Lam}.
 We say that $\Lambda$ is $(\alpha,d)-$Diophantine if its frequency $\omega(\xi)$, defined in
 (\ref{frequence}), satisfies
 \begin{equation}  \label{dn alpha-d dioph}
     \omega(\xi) \in D_{\alpha,d}= \big \{   \omega \ \in \mathbb R^2 \big |  \   | \langle \omega,k \rangle |  \geq \frac{\alpha}{ |k|^{1+d}}, \
     \forall \ k \in \mathbb Z^2 \backslash \{ 0 \} \big \}.
 \end{equation}
If \eqref{dn alpha-d dioph} holds for some $\alpha >0 $, and $d>0$, we say that the torus
$\Lambda$ (also its frequency) is uniformly Diophantine.
\end{defi}
Note also that when $d>0$ is fixed, the Diophantine property (for some $\alpha >0$) of $\Lambda$ is independent of the selected action-angle coordinates, see \cite{Broer07}. If $\Lambda$ is $(\alpha,d)-$Diophantine, then its frequency must be irrational.

It is known that the set $D_{\alpha,d} $ is a closed set with closed half-line structure. When we take
$\alpha$ to be sufficiently small, it is a nowhere dense set but with no isolated points, and its
measure tends to full measure as $\alpha$ tends to $0$: the measure of its complement is of order
$\mathcal{O}(\alpha)$. On the other hand, the trace of $D_{\alpha,d} $ on the unit sphere is a
Cantor set. See Refs. \cite{Poschel01}, \cite{HB90}, and \cite{uni-kam}.

\begin{defi}  \label{dn bonnes valeurs}
        For some $\alpha>0 $ and some $d>0$, we define the set of good values associated with a energy level $E$, denoted by $\mathcal{G}(\alpha,d, E)$, obtained from $\cup_{a \in J}Q_\infty(a)$ by removing the following set of bad values $\mathcal{B}(\alpha,d, E)$:
\beas
\mathcal{B}(\alpha,d, E)& =  & \Bigg ( \bigcup_{dist(a,S) < \alpha} Q_\infty(a) \Bigg )
                                \bigcup \Bigg ( \bigcup_{a \in J\backslash S: \ \omega(\xi_a) \textrm{ is not }
                                    (\alpha,d)- \textrm{Diophantine}} Q_\infty(a)\Bigg )
                                          \\
                  &         &  \bigcup \Bigg( \bigcup_{a \in J\backslash S: \ |d \langle q \rangle(\xi_a) | < \alpha }
                                            Q_\infty(a) \Bigg )
 \bigcup \Bigg ( \bigcup_{a \in J\backslash S: \ |\omega'(\xi_a)| < \alpha } Q_\infty(a) \Bigg ) .
\eeas
\end{defi}

\begin{rema} \label{rem2}  % ~~\\
         \begin{enumerate}[(i)] \label{peti mesure}
         \item When $d>0$ is kept fixed, the measure of the set of bad values
             $\mathcal{B}(\alpha,d, E)$ in $\cup_{a \in J}Q_\infty(a)$ (and in $\langle q
             \rangle_{\Lambda_a} (J) $) is small together with $\alpha$, is $\mathcal O
             (\alpha)$, when $\alpha >0$ is small enough, provided that the measure of
                                    \begin{equation}  \label{a condition}
                                            \Bigg ( \bigcup_{a \in
                                             J\backslash S: \ \omega(\xi_a) \in \mathbb Q } Q_\infty(a)
                                             \Bigg ) \bigcup \Bigg ( \bigcup_{a \in S} Q_\infty(a)
                                            \Bigg )
                                    \end{equation}
             is sufficiently small, depending on $\alpha$ (see Ref. \cite{Hitrik07}).
             \item Let $G \in \mathcal{G}(\alpha,d, E) $ be a good value, then by definition of
                 $\mathcal{B}(\alpha,d, E) $ and remark (\ref{rem1}), there are a finite number
                 of corresponding $(\alpha, d)-$Diophantine tori
                 $\Lambda_{a_1},\ldots,\Lambda_{a_L}$, with $L \in \mathbb N^*$ and $\{
                 a_1,\ldots,a_L\} \subset J\setminus S $, in the energy space $p^{-1}(E) \cap
                 T^*M$, such that the pre-image
                                         $$\langle q \rangle ^{-1}(G)= \{\Lambda_{a_1},\ldots, \Lambda_{a_L} \}.$$
             In this way, when $G $ varies in $\mathcal{G}(\alpha,d, E) $, we obtain a Cantor
                 family of $(\alpha, d)-$Diophantine invariant tori  in the phase space
                 satisfying $ \{p= E, \langle q \rangle = G \}$.
         \end{enumerate}
\end{rema}

When $G \in \mathcal{G}(\alpha,d, E) $ is a good value, we define in the horizontal band of size
$\varepsilon$ of complex plan, given in \eqref{loca. spectre 2}, a suitable window of size
$\mathcal{O}(h^\delta) \times \mathcal{O}(\varepsilon h^\delta)$, around the \textit{good center $E+ iG$},
called \textit{good rectangle},
 \begin{equation} \label{cua so}
                        R^{(E, G)}(\varepsilon,h)
                                    = (E+i \varepsilon \ G)+  \Big[-\frac{h^\delta}{\mathcal{O}(1)},\frac{h^\delta}{\mathcal{O}(1)} \Big]
                                    +i \varepsilon \Big [ -\frac{h^\delta}{\mathcal{O}(1)}, + \frac{h^\delta}{\mathcal{O}(1)} \Big ].
 \end{equation}

Now let $G \in  \mathcal{G}(\alpha,d, E)$ be a good value. As in Remark \ref{rem2}, there exists
$L$ elements in pre-image of $G$ by $\langle q \rangle$. The spectral results related to each of these elements are similar.
Therefore, to simplify to announce the results (in Theorem \ref{theorem quasi-spectre}), we shall assume that $L=1$ and we write
    \begin{equation} \label{preimage}
     \langle q \rangle ^{-1}(G)=\Lambda_ a \subset  p^{-1}(E) \cap T^*M , \ a \in J\setminus S ,
    \end{equation}
Note that this hypothesis can be achieved if we assume that the function $\langle q \rangle$ is
injective on $J\setminus S $.

\begin{defi}[Refs. \cite{lectureColin}, \cite{Arnold67}, and \cite{Cappel94}]   \label{ind}
% xem dinh ngia grassmanienne pag 16 quyen Yves.
Let $ E $ be a symplectic space and let $ \Lambda (E) $ be his Lagrangian Grassmannian  (which is
set of all Lagrangian subspaces of $ E $). We consider a bundle $ B $ in $ E $ over the circle or a
compact interval provided with a Lagrangian subbundle called vertical. Let $ \lambda (t) $ be a
section of $ \Lambda (B) $ which is transverse to the vertical edges of the interval in the case
where the base is an interval. The Maslov index of $ \lambda (t) $ is the intersection number of
this curve with the singular cycle of Lagrangians which do not cut transversely the vertical
subbundle.
\end{defi}

Let $\Lambda_a$, $ a  \in J\backslash S$ be an invariant Lagrangian torus and let $\kappa$ be
the action-angle local coordinates in \eqref{coor}. The fundamental cycles $(\gamma_1,\gamma_2)$ of
$\Lambda_a$ are defined by
            $$ \gamma_j= \kappa^{-1} ( \{ (x, \xi) \in T^* \mathbb{T}^2: x_j=0, \xi= \xi_a \}) , \ j=1,2. $$
Then we note $\eta \in \mathbb Z^2$ the Maslov index and $S \in \mathbb R^2$ the integral action of
these fundamental cycles,
\begin{equation} \label{act}
    S= (S_1, S_2) = \left( \int_{\gamma_1} \theta , \int_{\gamma_2} \theta  \right),
\end{equation}
where $\theta$ is locally a (primitive) Liouville $1-$ form of the closed form $\sigma$ on
$(\Lambda_{a}, T^*M)$, whose existence is ensured by the Poincare lemma.

% Dua vao loi dan o day: quelques expression ici? We treat the stand case with E=0.

We recall here the result of asymptotic spectrum treated for the stand case at the energy level
$E=0$, given by \cite{Hitrik07}. However this result can be immediately generalized for any energy
lever $E$ by a translation, that we will carry out further in Sec. \ref{sp-lattice Sec}.

\begin{theo}[\cite{Hitrik07}]  \label{theorem quasi-spectre}
    For $E=0$ and assume that action-angle coordinates $\kappa$ in \eqref{coor} send $\Lambda_a$ to the zero
    section $ \mathbb T^2 \times \{\xi_a=0\} \in T^* \mathbb T^2$.
    Suppose that $P_\varepsilon$ is an operator with principal symbol (\ref{symb prin}) which satisfies
    Assumptions \ref{hypothese}.
    Assume that $h \ll \varepsilon = \mathcal{O}(h^\delta)$ for $0< \delta <1 $.
  Let $G \in  \mathcal{G}(\alpha,d, 0)$ be a good value as Defenition \ref{dn bonnes valeurs}, and assume that \eqref{preimage} is true. Then the eigenvalues $\mu$ of $ P_ \varepsilon $
  with multiplicity in the good rectangle $  R^{(0, G)}(\varepsilon,h)$ of the form \eqref{cua so} have the following
  expression:
  \begin{equation} \label{eigenvalues}
  \mu= P ^{(\infty)} \Big( h(k-\frac{\eta}{4})-\frac{S}{2 \pi};\varepsilon, h\Big) + \mathcal O(h^\infty),
                                        k \in \mathbb Z^2,
  \end{equation}
    where $P^{(\infty)}(\xi; \varepsilon, h )$ is a smooth function of $\xi$ evolving in a neighborhood of $(0, \mathbb R^2)$
    and $\varepsilon, h$ in neighborhoods of $(0, \mathbb R)$.
    Moreover $P^{(\infty)}(\xi; \varepsilon, h )$ is real valued for $\varepsilon =0$ and it admits the following polynomial asymptotic
    expansion in $(\xi, \varepsilon, h)$ for the $C^\infty-$ topology:
    \begin{equation} \label{symbole normal}
        P^{(\infty)} ( \xi; \varepsilon, h) \sim    \sum_{\alpha, j,k} C_{\alpha j k} \ \xi^\alpha \varepsilon^ j h^k,
     \end{equation}
    Particularly $P^{(\infty)}$ is classical in the space of symbols with $h-$leading term:
    \begin{equation} \label{prin normal}
            p_{0}^{(\infty)} (\xi, \varepsilon)= p(\xi)+ i \varepsilon \langle q \rangle (\xi) +
                                             \mathcal O(\varepsilon ^2).
    \end{equation}
    Here $p, q$ are the expressions of $p,q$ in action-angle variables near $\Lambda_a$, given by (\ref{coor}), and  $\langle q \rangle $
    is the average of $q$ on tori, given in (\ref{moyenne2}).
\end{theo}

\begin{rema} \label{diffeo}
We can write the total symbol $ P^{(\infty)}$ in the reduce form:
     \begin{equation}
         P^{(\infty)} ( \xi; \varepsilon, h) =  p(\xi)+ i \varepsilon \langle q \rangle (\xi) +   \mathcal O(\varepsilon ^2)+ \mathcal O(h),
    \end{equation}
uniformly for $\varepsilon$ and $h$ small. Note that $dp$, $d \langle q \rangle $ are linearly
independent in $\xi=0$, and in the regime $h \ll \varepsilon$ we can show that the function
$P^{(\infty)}$ is a local diffeomorphism from a neighborhood $B(0, r)$, $r>0 $, of
    $\xi=0 \in \mathbb R^2$ into its image, which is in a $\mathcal O(\varepsilon)-$ horizontal band of the complex plane and covers the good rectangle (see Ref. \cite{QS14}).
\end{rema}

\begin{rema}
By Theorem \ref{theorem quasi-spectre}, it is true that the eigenvalues of $P_\varepsilon$ in the good rectangle $  R^{(0, G)}(\varepsilon,h)$
form a deformed lattice. In the case  $L\geq 1$, then the spectrum of $P_{\varepsilon}$ in $  R^{(0, G)}(\varepsilon,h)$ are union of $L$ such deformed lattices.
\end{rema}

% Nahn xet truong hop dac biet khi p,q commute thi ko co term epsilon mu 2.

%%%%%%%%%%%%%%%%%%%%%%%%%%%%%%%%%%%%%%%%%%
%%%%%%%%%%%%%%%%%%%%%%%%%%%%%%%%%%%%%%%%%%   TRUONG HOP EVERY E

\subsection{Spectral asymptotic pseudo-lattice and its monodromy}   \label{sp-lattice Sec}

% Chung ta se thiet lap 1 xd de chung to rang spectre la mot asymptotic pseudo lattice. Tu do ta co the ap dung monodomy tu article...
% Su khac biet va kho khan o day la j? sai khac epsilon la dang ke trong spectral band. Ta se xem xet ca epsilon coi nhu 1 classical chu ko chi h.
% Trong article kia da co san mot he integrable, o day ta phai xd no, cong thuc ve spectre cung khac.
% Tim ra nhung bat bien ko phu thuoc vao h, epsilon???

We are considering the operator $P_\varepsilon$ with all assumptions as in above Sec. \ref{s a}.
In this section, we first apply the result of the stand case, stated in Theorem \ref{theorem
quasi-spectre}, to given an asymptotic expansion of the eigenvalues of $P_\varepsilon$ in any good
rectangle $R^{(E, G)}(\varepsilon,h)$ of the spectral band. After, we will give a brief construction to
show that the spectrum satisfies all conditions of an asymptotic pseudo-lattice. We refer to \cite{QS14} for a similar construction with more
detail.

We note that $\langle q \rangle = \langle q \rangle_{\Lambda}$, the average of $q$ over the regular $H_p-$invariant Lagrangian
torus $\Lambda$, given in \eqref{moyenne de q}, can be seen as a constant function on
$\Lambda$. In this way $\langle q \rangle$ well defines a analytic function on the union of the
regular invariant Lagrangian tori. It is true that the fibration of all the regular invariant Lagrangian tori $\Lambda$, given by
the momentum map $F$, fills almost completely the phase space $T^*M$.
% Theo dinh ly Serre:  tap cac diem ky di co do do Lebesgue bang 0.
So we assume that the function $\langle q \rangle$ can be extended smoothly on the whole phase
space $T^*M$. Moreover, we assume that the differentials $d p$ and $d \langle q \rangle$ are
$\mathbb R-$linearly independent almost everywhere on $T^*M$. It is clear that $p$ and $\langle q \rangle$
commute in neighborhood of each regular torus $\Lambda_a$. So we can see $(p, \langle q \rangle)$
as a completely integrable system and we take now the momentum map $F=(p, f)$ in \eqref{F}, with $f= \langle q \rangle$.

We assume moreover that the map $F$ is proper and has connected fibers. This ensures that all regular
fibres of $F$ are invariant Lagrangian tori-the Liouville tori.
Denote $U_r$ the set of all regular values of $F$ and let $U$ be a open subset of $U_r$ with compact
closure.

We assume further that all assumptions of Sec. \ref{hypothese} are true for any energy level $E$, taken in a
bounded interval of $\mathbb R$.
Of course, we can take $E$ in the interval $p_x(U)$, where $p_x$ is the projection on the real axe of variable
$x$.

Let a point $c \in U$. Then by the angle-action Theorem \ref{A-A}, we have local angle-action variables on a neighborhood of the torus $\Lambda_c= F^{-1}(c)$ in $T^*M$ as in \eqref{coor}: there exists a sufficiently small open neighborhood $ U^c \subset U$ of $c$, a symplectomorphism
$\kappa= (x,\xi): V \rightarrow \mathbb T ^n \times  A$, here $V:=F^{-1}( U^c) $ and $A  \subset \mathbb  R^2$ is a small open set,
a diffeomorphism $\varphi= (\varphi_1, \varphi_2): A \rightarrow U^c$ such that $F \circ \kappa ^{-1}(x, \xi)= \varphi (\xi)$, for all $ x \in \mathbb T^2, \ \xi \in A $.

Therefore, as in \eqref{p} and \eqref{moyenne2}, the expressions of the functions $p$ and $\langle q \rangle$ in the above action-angle coordinates is dependent only on action variables,
$$p(\xi)=\varphi_1 (\xi),   \langle q \rangle(\xi) = \varphi_2 (\xi), \ \xi \in A. $$
The fact that the horizontal spectral band is of size $\mathcal{O}(\varepsilon)$ suggests us
introducing the function
        \begin{eqnarray}   \label{chi}
            \chi :  \mathbb R^2 \ni u= (u_1,u_2) \mapsto \chi_u &=& (u_1, \varepsilon u_2) \in \mathbb R^2
            \\
            &\cong & u_1+i \varepsilon u_2 \in \mathbb C,     \nonumber
        \end{eqnarray}
in which we identify $\mathbb C$ with $\mathbb R^2$.
We denote
\beq \label{elp}
U^c ( \varepsilon )=  \chi( U^c), \ U(\varepsilon)=  \chi(U).
\eeq

Now, with any point $a=(E,G) \in U^c $ such that $G$ is a good value, i.e., $G \in \mathcal{G}(\alpha,d,
E)$, as in Definition \ref{dn bonnes valeurs}. We notice also that with assumptions of $F$, the condition \eqref{preimage} is valid ($L=1$).
Then the corresponding torus $\Lambda_a= F^{-1}(a)$ is $(\alpha,d)-$Diophantine, as discussed in Remark \ref{rem2}. We suppose that by $\kappa$, $\Lambda_a \simeq \Lambda_{\xi_a}$, with $\xi_a= \varphi^{-1}(a) \in A$.
Let $\eta \in \mathbb Z^2$ be the Maslov index, and $S \in \mathbb R^2$ be the integral action of
the fundamental cycles of $\Lambda_a$.

Noticing that $ \sigma(P_\varepsilon)= \sigma(P_\varepsilon - \chi_a  ) +\chi_a$ and applying Theorem \ref{theorem quasi-spectre}
(in the stand case) for the operator $(P_\varepsilon - \chi_a)$ with respect to the good rectangle $R^{(0, G)}(\varepsilon,h)$, we obtain easily the asymptotic eigenvalues of $P_\varepsilon$ in the good rectangle $R^{(E, G)}(\varepsilon,h)$, as following.

All eigenvalues $\mu$ of $P_\varepsilon$ in the good rectangle $R^{(E,
G)}(\varepsilon,h)$, defined by \eqref{cua so}, with modulo $\mathcal O(h^\infty)$, are \textit{micro-locally} given by
 \begin{equation} \label{eigenvalues-E}
   \sigma(P_\varepsilon) \cap R^{(E, G)}(\varepsilon,h) \ni \mu=
   P \Big( \xi_a+ h(k-\frac{\eta}{4})-\frac{S}{2 \pi};\varepsilon, h\Big) + \mathcal O(h^\infty),
                                        k \in \mathbb Z^2,
  \end{equation}
where $P(\xi; \varepsilon, h)$ (for ease the notation, writing
$P$ instead of $P ^{(\infty)}$, given in Theorem \ref{theorem quasi-spectre}) is a smooth function of $\xi$ in a
neighborhood of $(\xi_a, \mathbb R^2)$ and of $\varepsilon, h$ in neighborhoods of $(0, \mathbb R)$.
Moreover, $P(\xi; \varepsilon, h )$ admits an asymptotic expansion in $(\xi, \varepsilon, h)$ of the form \eqref{symbole normal}, and in particularity the $h-$leading term of $P$ is of the form \eqref{prin normal}:
\bea \label{prin normal-E}
            p_{0}(\xi, \varepsilon) & = &p(\xi)+ i \varepsilon \langle q \rangle (\xi) +
                                             \mathcal O(\varepsilon ^2)
            \\
            &=&      \varphi_1 (\xi) + i \eps \varphi_2 (\xi) + \mathcal O(\varepsilon ^2) .
\eea

\begin{rema} ~\\
\begin{enumerate}[(i)]
\item The previous result shows that the eigenvalues of $P_\varepsilon$ in a good rectangle
form a deformed \textit{micro-lattice}, it is image of a square lattice of $h \mathbb Z^2$ by a
local diffeomorphism. Moreover, we can show that the lattice has a horizontal spacing $ \mathcal O(h)$
and a vertical spacing $ \mathcal O(\varepsilon h)$, see \cite{QS14}.
\item In this paper, the vocabulary \textit{micro} is used for some objects, which are related to a very small domain depending on $h$.
\end{enumerate}
\end{rema}

According the definition of an \textit{asymptotic pseudo-lattice}, given in \cite{QS14}, which is recalled in Definition \ref{pseu-lattice}, and with the help of the micro-locally asymptotic spectrum \eqref{eigenvalues-E}, the next construction is to show the following statement.
% upcoming
\begin{theo}
The spectrum of $P_\varepsilon$ on the domain $U(\varepsilon)$ is an asymptotic pseudo-lattice.
\end{theo}

As a classical property, it is true that the difference between the integral actions,
with factor $\frac{1}{2 \pi}$, and the action coordinates are locally constant on every regular tori
$\Lambda_a \subset V$ (see Ref. \cite{QS14}):
 \begin{equation}  \label{pt action}
                             \frac{S}{2 \pi}- \xi_a := \tau_c \in \mathbb R^2,
 \end{equation}
 is locally constant in $c \in U_r$.

From Remark \ref{diffeo}, we know that Eq. \eqref{eigenvalues-E} provides a bijective
correspondence between the eigenvalues in the good rectangle $R^{(E, G)}(\varepsilon,h)$ and $hk$
in a part of $h \mathbb Z^2$. Moreover, this correspondence is given by a smooth local
diffeomorphism in $E \in \mathbb C$, denoted by
 $f= f(\mu; \varepsilon, h)$, which sends micro-locally $R^{(E, G)}(\varepsilon,h)$ to its image, denoted by $E(a,
 \varepsilon,h)$ (which is close to $\frac{S}{2 \pi}$), such that $\mu \in \sigma(P_\varepsilon) \cap R^{(E, G)}(\varepsilon,h)$ is sent
 to $hk \in h \mathbb Z^2 $, modulo $\mathcal O(h^\infty)$:
 \begin{eqnarray}  \label{new hk}
                 f= f(\mu, \varepsilon;h) & =& \tau_c + h \frac{\eta}{4}+ P^{-1}(\mu)
                  \\
                      f (R^{(E, G)}(\varepsilon,h)) &= & E(a,\varepsilon,h)
                       \nonumber \\
                \sigma(P_\varepsilon) \cap R^{(E, G)}(\varepsilon,h) \ni  \mu & \mapsto &
                        f(\mu,\varepsilon; h) \in h \mathbb Z^2 +\mathcal O(h^\infty).  \nonumber
 \end{eqnarray}

We say that the map $f$ in \eqref{new hk} is a \textit{micro-chart} of the spectrum of $P_\varepsilon$ on
the good rectangle $R^{(E, G)}(\varepsilon,h)$. We shall analyze it.

% Chung ta se analysis ham  nay:

Let $\widetilde{f}= f \circ \chi$, then we have
\begin{equation}    \label{f tilde}
\widetilde{f}=  \tau_c + h \frac{\eta}{4}+ P^{-1} \circ \chi .
\end{equation}

To analysis $P^{-1} \circ \chi$, we first discuss about its inverse, $\widehat{P}:= \chi^{-1} \circ
P$. It is obtained from $P$ by dividing the imaginary part of $P$ by $\varepsilon$. As $P$ admits
an asymptotic expansion in $(\xi, \varepsilon,h)$, so it is true that $\widehat{P}$ admits an
asymptotic expansion in $(\xi, \varepsilon,\frac{h}{\varepsilon})$ (here $h \ll \varepsilon $).
Moreover, we can write $\widehat{P}$ in the reduce form as below:
\begin{eqnarray}
\widehat{P}(\xi, \varepsilon,h) & = &  \widehat{P}_0(\xi)+ \mathcal O(\varepsilon) + \mathcal O(\frac{h}{\varepsilon})
                       \nonumber \\
                            & = & \widehat{P}_0(\xi) + \mathcal O(\varepsilon, \frac{h}{\varepsilon}),
\end{eqnarray}
uniformly for $h, \varepsilon $ small and $h \ll \varepsilon $, with $\widehat{P}_0(\xi)=
\varphi_1(\xi)+ i \varphi_2(\xi) $.

 \begin{prop}(see \cite{QS14})  \label{D.A inverse}
            Let $\widehat{P}= \widehat{P}(\xi; X)$ a complex-valued smooth function of $\xi$
            near $0 \in \mathbb R^2$ and $X$ near $0 \in \mathbb R^n$.
            Assume that $\widehat{P}$ admits an asymptotic expansion in $X$ near $0$ of the form
                $$\widehat{P}(\xi; X) \sim  \sum_\alpha C_\alpha(\xi) X^\alpha,$$
            with $C_\alpha(\xi)$ are smooth functions and $C_0(\xi):=\widehat{P}_0(\xi) $ is local diffeomorphism near $\xi=0$.

            Then, for $\mid X \mid $ small enough, $\widehat{P}$ is also a smooth local diffeomorphism near $\xi=0$ and its inverse admits an asymptotic expansion in $X$ near $0 $ whose the first term is
            $(\widehat{P}_0)^{-1}$.
\end{prop}

This proposition ensures that the map $\widehat{P}^{-1}= P^{-1} \circ \chi$ admits an asymptotic expansion in $(\varepsilon,\frac{h}{\varepsilon})$ whose first term is $(\widehat{P}_0)^{-1}=  \varphi ^{-1}$. So from \eqref{f tilde}, $\widetilde{f}$ admits an asymptotic expansion in
$(\varepsilon, \frac{h}{\varepsilon})$ with the leading term
            \begin{equation}  \label{pre terme f tilde}
                 \widetilde{f}_0= \tau_c+  \varphi^{-1}.
            \end{equation}
We notice that the leading term $\widetilde{f}_0$ is a local diffeomorphism, completely defined on
$U^c$. It does n't depend on the selected good rectangle, is valid for any good value $a \in
U^c$.

However, in the domain $U^c(\varepsilon )$, we have a lot of good rectangles. They could be disjoint when $h$ is sufficiently small, and despite their density, not quite fill $U^c(\varepsilon )$. In fact we know moreover
from Remark \ref{rem2} that the set of good values is a Cantor set, and outside a set of small
measure. We have therefore locally a Cantor family of micro-charts with common leading term that is well locally defined. This family is
called a \textit{spectral pseudo-chart} on the domain $U^c(\varepsilon )$.

The above construction ensures that the spectrum of $P_\varepsilon$ on the domain $U(\varepsilon)$
satisfies all conditions of a particular discrete lattice-an asymptotic pseudo-lattice, given in
\cite{QS14}, that we recall here:

\begin{defi} \label{pseu-lattice}
     Let $U$ be a open subset of $\mathbb R^2$ with compact closure. We denote $U(\varepsilon)$ as in \eqref{elp}, and
     let $\Sigma(\varepsilon, h)$ (which depends on small $h$ and $\varepsilon$) be a discrete subset of $U(\varepsilon)$.
     For $h, \varepsilon$ small enough and in the regime $ h \ll \varepsilon$, we say that $(\Sigma(\varepsilon, h), U(\varepsilon))$
     is an asymptotic pseudo-lattice if:
         for any small parameter $\alpha >0$, there exists a set of good values in $\mathbb R^2$, denoted by $\mathcal{G}(\alpha)$, whose complement is of small measure together with $\alpha$ in the sense:
             $$\mid {}^C \mathcal{G} (\alpha) \cap I  \mid \leq C \alpha \mid I \mid, $$
with a constant $C>0$ for any domain $I \subset \mathbb R^2$.  \\
For all $c \in U$, there exists a small open subset $U^c \subset U $ around $c$ such that for every good value $a \in U^c  \cap \mathcal{G}(\alpha)$, there is a adapted good rectangle $R^{(a)}(\varepsilon,h) \subset U^c( \varepsilon)$ of the form \eqref{cua so} (with $a=(E,G)$), and a smooth local diffeomorphism $f= f(\cdot; \varepsilon,h)$ which sends $R^{(a)}(\varepsilon,h)$ on its image, satisfying
\beq  \label{semi-cart}
    \Sigma( \varepsilon, h) \cap R^{(a)}(\varepsilon,h) \ni  \mu \mapsto  f(\mu;\varepsilon, h) \in h \mathbb Z^2 +\mathcal O(h^\infty).                        \eeq
Moreover, the function $\widetilde{f}:= f \circ \chi$, with $\chi$ defined by \eqref{chi}, admits asymptotic expansions in
$(\varepsilon,\frac{h}{\varepsilon})$ for the $C ^\infty-$ topology in a neighborhood of $a$,
uniformly with respect to the parameters $h$ and $\varepsilon$, such that its leading term $\widetilde{f}_0$
is a diffeomorphism, independent of $\alpha$, locally defined on the whole $U^c $ and independent
of the selected good values $a \in U^c $.

We also say that the couple $(f(\cdot; \varepsilon,h), R^{(a)}(\varepsilon,h) )$ is a micro-chart,
and the family of micro-charts $(f(\cdot; \varepsilon,h), R^{(a)}(\varepsilon,h) )$, with all $a
\in U^c \cap \mathcal{G}(\alpha) $, is a local pseudo-chart on $U^c( \varepsilon)$ of
$(\Sigma(\varepsilon, h), U(\varepsilon))$.
\end{defi}

\begin{rema}
% There exists an other lattice with some similar properties but lighter, called asymptotic lattice, given in \cite{Vu-Ngoc99} to define the quantum monodromy. The last lattice is locally defined, while the asymptotic pseudo-lattice is very fine, it is micro-locally defined.
The introduction of this discrete lattice aims to show that the monodromy that we will
define is directly built from the spectrum of operators. If different operators have the same
spectrum, then they have the same monodromy.
\end{rema}

Now we can define the spectral monodromy of the operator $P_\varepsilon$ on the domain $U
(\varepsilon) $, denoted by $[\mathcal M _{sp}]$, as the monodromy of the asymptotic pseudo-lattice
$(\sigma(P_\varepsilon), U(\varepsilon) )$, as the following.

Let $\{ U^j \}_{j \in \mathcal{J} }$, here $\mathcal{J}$ is a finite index set, be an arbitrary (small enough) locally finite covering of $U
$. Then the asymptotic pseudo-lattice $(\Sigma(\varepsilon, h), U(\varepsilon))$ is covered by the associated local pseudo-charts $ \{ \left(  f_ j (\cdot; \varepsilon, h),  U^j (\varepsilon) \right) \}_{j \in \mathcal{J} }$, here $ U^j  (\varepsilon)= \chi( U^j)$. Note from Definition \ref{pseu-lattice} that the leading terms $ \widetilde{f}_{j,0}(\cdot; \varepsilon, h)$ are well defined on whole $ U^j$ and we can see them as
the charts of $U$. Analyzing transition maps, we have the following result.
\begin{theo} [\cite{QS14}]
 On each nonempty intersection $  U^i \cap U^j \neq \emptyset$, $i, j \in \mathcal{J}$, there exists a unique integer linear map $M_{ij} \in GL(2,
\mathbb Z)$ (independent of $h, \varepsilon$) such that:
\begin{equation} \label{transition-pseu}
d \big ( \widetilde{f}_{i, 0} \circ (\widetilde{f}_{j, 0})^{-1} \big )= M_{ij}.
\end{equation}
\end{theo}

Then we define the class, denoted by $[\mathcal M _{sp}] \in \check{H}^1(U,GL(2, \mathbb Z) )$, obtained from the $1$-cocycle of $\{M_{ij} \} $ with modulo coboundary, in the \v{C}ech cohomology of $U$ with values in the integer linear group $GL(2, \mathbb Z)$. It is called the (linear) monodromy of the asymptotic pseudo-lattice $(\Sigma(\varepsilon, h), U(\varepsilon))$.
It does n't depend on the selected finite covering $\{ U^j \}_{j \in \mathcal{J} }$.

We can also associate the class $[\mathcal M _{sp}]$ with its holonomy, that is a group morphism from the fundamental group $\pi_1(U)$ of $U$ to the group $ GL(2, \mathbb Z)$, modulo conjugation. Their trivial property is equivalent.

% For more detail of this point, we refer to Ref. \cite{QS14}.

\begin{defi} \label{monodromie spectrale}
For $\varepsilon, h > 0$ sufficiently small such that $h \ll \varepsilon \leq h^\delta, 0< \delta< 1$, the
spectral monodromy of the operator $P_\varepsilon$ on the domain $U(\varepsilon)$,
is the class $[\mathcal M _{sp}] \in \check{H}^1(U,GL(2, \mathbb Z) )$.
\end{defi}

%%%%%%%%%%%%%%%%%%%%%%%%%%%%%%%%%%%%%%%%%%%%%%%%%%%%%%%%%%%%%%%%%%%%%
%%%%%%%%%%%%%%%%%%%%%%%%%%%%%%%%%%%%%%%%%%%%%%%%%%%%%%%%%%%%%%%%%%%%%

\subsection{Spectral monodromy recovers the classical monodromy}

% Toan tu dc xet la related toi 1 he kha tich. Monod class cua no la dc biet tu lau. Ta dua ra moi lien he giua cac monod.
% Them Refs...

Our considered operators $P_\varepsilon$ in the first case  are related to integrable classical systems. With the property of classical integrability, a geometrical invariant of such systems that obstructs the existence of global angle-action variables on the phase space, is known (see Refs. \cite{Mineur48}, \cite{Audin01}, and \cite{Cushman97}). That is the classical monodromy, defined correctly by Duistermaat \cite{Duis80}.
We make here a relationship between the spectral monodromy and the classical monodromy.

\begin{defi}
 A completely integrable system on a symplectic manifold $(W, \sigma)$ of dimension $ 2n $ ($n \geq 1$) is given $n$ smooth real-valued functions $f_1, \dots, f_n$ in involution with respect to the Poisson bracket generated from the symplectic form $ \sigma$, whose differentials are almost everywhere linearly independent.
In this case, the map $$ F=(f_1, \dots, f_n ): M \rightarrow \mathbb R^n $$ is called \textit{momentum map or completely integrable system}.
\end{defi}

Let $U$ be a open subset of the set of regular values of $F$. Then we have:

\begin{theo}[Angle-action theorem]  \label{A-A}
Let $c \in U$, and $\Lambda_c$ be a compact regular leaf of the fiber $F^{-1}(c)$. Then there exists an open neighborhood $V= V^c$ of $\Lambda_c$ in $W$ such that $F\mid_ {V} $ defines a smooth locally trivial fibre bundle onto an open neighborhood $ U^c \subset U$ of $c$, whose fibres are invariant Lagrangian $n-$tori. Moreover, there exists a symplectic diffeomorphism $\kappa = \kappa^c$,
    $$\kappa= (x,\xi): V \rightarrow \mathbb T ^n \times  A, $$
with $A= A^c \subset \mathbb R^n$ is an open subset, such that $F\circ \kappa^{-1}(x, \xi)= \varphi(\xi)$ for all $x \in \mathbb T^n $,
and $\xi \in A$, and here $\varphi= \varphi^c: A \rightarrow \varphi (A)=U^c $ is a local diffeomorphism. We call $( x, \xi)$ local angle-action variables near $\Lambda_c$ and $(V, \kappa)$ an local angle-action chart.
\end{theo}
% $x=(x_1, \dots, x_n) \in \mathbb T^n $,  and $\xi= ( \xi _1, \dots \xi_ n ) \in A$,

Note that one chooses usually the local chart such that the torus $\Lambda_c$ is sent by $\kappa$ to the zero section $T ^n \times \{0 \}$. By this theorem, for every $a \in U^c $, then $\Lambda_a:= F^{-1}(a) \cap V^c$ are invariant Lagrangian $n-$tori, called Liouville tori, and we have $\Lambda_a \simeq \Lambda_{\xi_a}$, with some $\xi_a \in A$.

% its image by $\kappa^c$ is a $n-$torus $T ^n \times \{\xi_a \}$, denoted by $\Lambda_{\xi_a}$, . We write often

Moreover, there exists an affine structure on the geometry of Liouville tori (see Refs. \cite{Duis80}, and \cite{Vu-Ngoc06}). We assume for simplicity that $F$ is proper and connected fibres. Then every regular fibre $\Lambda_c:= F^{-1}(c)$, with $c \in U$, is a Liouville torus.

When $U$ is assumed furthermore relatively compact subset and we cover it by an arbitrary small enough finite open
covering $\{ U^j \}_{j \in \mathcal{J} }$, here $\mathcal{J}$ is a finite index set. Then the relatively
compact subspace $X= F^{-1}(U)$ is also covered by a finite covering of angle-action charts $ \{
(V^j, \kappa ^j ) \} _ {j \in \mathcal{J} }$. Here $V^j = F^{-1} (U^j)$, $ j \in \mathcal{J}$.
It is classical that on the nonempty overlaps $V^i \cap V^j$, there is a locally constant matrix $M _{ij}^{cl} \in GL(2, \mathbb Z)$, and a constant $ C_{ij} \in \mathbb R^2$, with $i, j \in
\mathcal{J}$, such that
        \begin{equation} \label{affine}
        ({\varphi ^i}) ^{-1} \circ \varphi ^j (\cdot)  =  M _{ij}^{cl}(\cdot)+ C_{ij}.
        \end{equation}
The non-triviality of $M _{ij}$ leads surely to a obstruction of the existence of
global action-angle coordinates, called classical monodromy. It
is defined as the $\mathbb Z^n-$ bundle $H_1(\Lambda_c, \mathbb Z) \rightarrow c \in U $, where
$H_1(\Lambda_c, \mathbb Z)$ is the first homology group of $\Lambda_c$, whose transition maps
between trivializations are
$$\{ {}^t \big ( d( ({\varphi ^i}) ^{-1} \circ \varphi ^j  ) \big ) ^{-1}=  {}^t \big ( M_{ij}^{cl} \big ) ^{-1} \}.$$
The triviality of this monodromy is equivalent to the existence of global action variables on the
space of Liouville tori. For more on this monodromy, we refer also to Ref. \cite{Vu-Ngoc06}.

With $W= T^*M$ and $n=2$, we apply now the result \eqref{affine} to spectral pseudo-charts of the operator
$P_\varepsilon$, and noticing the expression \eqref{pre terme f tilde}. It is true that on the nonempty intersections
$U^i \cap U^j \neq \emptyset$, the corresponding local spectral pseudo-charts of $(\sigma(P_\varepsilon),
U(\varepsilon) )$ satisfy
\begin{equation}
d \big ( \widetilde{f}_{i, 0} \circ (\widetilde{f}_{j, 0})^{-1} \big )=  M _{ij}^{cl} \ \in GL(2, \mathbb Z).
\end{equation}
This previous result, due to classical results, is independently found again from one of asymptotic pseudo-lattices, see \eqref{transition-pseu} cited from \cite{QS14}.

So we can state that in dimension $2$, the classical monodromy of an integrable system is completely identified with the spectral monodromy of small
non-selfadjoint perturbations of selfadjoint classical operators, accepting this integrable system as the leading term of the unperturbed part. And moreover:
\begin{theo}
The spectral monodromy of $P_\varepsilon$ is the adjoint of the classical monodromy.
\end{theo}
This result is the same one of paper \cite{QS14}, but proposed operators here are in a more general context.

%%%%%%%%%%%%%%%%%%%%%%%%%%%%%%%%%%%%%%%%%%%%%%%%%%%%%%%%%%%%%%%%%%%%%%%%%%%%%%%%
%%%%%%%%%%%%%%%%%%%%%%%%%%%%%%%%%%%%%%%%%%%%%%%%%%%%%%%%%%%%%%%%%%%%%%%%%%%%%%%%

\section{Spectral monodromy in the quasi-integrable case}

% Trang 2:  general assumption that the real energy space of the unperturbed leading symbol contains several flow invariant Lagrangian tori
% satisfying a Diophantine condition.

In this section we are studying the spectral monodromy of small non-selfadjoint perturbations of selfadjoint classical operators in $2$ dimension in the case when the leading symbol of the unperturbed part is quasi-integrable, together with a globally isoenergetic condition.

A radical general assumption given in \cite{Hitrik07}, in the spectral asymptotic construction of such operators from a $h-$ dependent complex window, is that the real energy surface at certain level of the unperturbed leading symbol possesses several Hamiltonian flow invariant Lagrangian tori, satisfying a uniformly Diophantine condition.

In the first case when the unperturbed principal symbol is completely integrable, this assumption is ensured as we known in Sec. \ref{s a}. However, it still satisfactory when this symbol is close to a completely integrable one, according the classical Kolmogorov-Arnold-Moser (KAM) theory (see Refs. \cite{Kol67}, \cite{Arn63}, \cite{Poschel01}, and \cite{HB90}).

We will give results about the asymptotic spectrum, similarly as in the first case, to show that the spectrum of such operators should be an asymptotic pseudo-lattice on the spectral band. Then we can define its spectral monodromy.

\subsection{Quasi-periodic flows of quasi-integrable systems}

Classical KAM theory allows to treat perturbations of a completely integrable Hamiltonian system. Under an isoenergetic condition, this theory proves the persistence of invariant Lagrangian tori, called KAM tori, on which the classical flow of the unperturbed system stills quasi-periodic with Diophantine constant frequencies. Moreover, the union of these KAM tori is a nowhere dense set, with complement of small measure in the phase space.

% To KAM, we refer to Refs.

We consider a perturbed Hamiltonian that is close to a completely integrable (non-degenerate) one:
\beq  p_\lambda= p+ \lambda p_1, \ 0< \lambda \ll 1,  \eeq where $p$ and $p_1$ are holomorphic
bounded Hamiltonian in a tubular neighborhood of $T^*M$, real on  $T^*M$  and furthermore $p$ is
assumed to be a completely integrable Hamiltonian system, as in Sec. \label{s a}.

Let $d >0$ fixed, and let $\Lambda_a$ be a $H_p-$invariant uniformly Diophantine Lagrangian torus in the energy space $p^{-1}(E)$ as Definition \ref{diop} with some $\alpha >0$.
In the angle-action coordinates $(x, \xi)$ on a neighborhood $V$ of $\Lambda_a$, given in \eqref{coor}, the function $p_\lambda$ becomes:
$$ p_\lambda \circ \kappa^{-1}= p_\lambda (x, \xi)=p (\xi)+ \lambda p_1 (x, \xi). $$
The Hamiltonian flow of $p$ on a $H_p-$invariant Lagrangian torus $\Lambda \subset V $ (close to $\Lambda_a$), $\Lambda \simeq \Lambda_\xi = \mathbb T^2 \times \{ \xi \}$, is quasi-periodic of constant Hamiltonian vector field \beq \label{Hp}
 H_p(x, \xi)=  \omega_1 (\xi)
\frac{\partial }{\partial x_1}+  \omega_2(\xi)  \frac{\partial }{\partial x_2}, \ x \in \mathbb
T^2, \ \xi \in A, \eeq with the frequency $\omega(\xi)$, given by \eqref{frequence}.

In particular, the frequency of $\Lambda_a \simeq  \Lambda_{\xi_a} $ is $\omega (\xi_a) =
\frac{\partial p}{\partial \xi} (\xi_a) $ satisfying the Diophantine condition \eqref{dn alpha-d
dioph}, for some $\alpha >0$.

\begin{defi} \label{Kol}
We say that $p$ is (Kolmogorov) local non-degenerate (on $V$) if the isoenergetic condition holds in the sense that the local
frequency map $\omega: A \rightarrow \mathbb R^2$, defined by \eqref{frequence}, is a
diffeomorphism onto its image.
\end{defi}
In fact, this condition is equivalent to
$$det  \frac{\partial \omega}{\partial \xi} = det (\frac{\partial^2 p}{\partial \xi^2}) \ \neq 0 \ \textrm{on} \ A,$$
and it means that $H_p-$invariant Lagrangian tori near $\Lambda_a$ can (locally) be parametrized by
their frequencies.

Let $\Omega= \omega(A) $ be the open range of the frequency map $\omega$.
% Let us recall that the set $D_{\alpha, d}$ given in \eqref{dn alpha-d dioph} denoted the set of $(\alpha,d)-$Diophantine frequencies.
Let $\Omega_{\alpha, d} \subset \Omega$ be the subset of frequencies which satisfy the Diophantine condition \eqref{dn alpha-d dioph} and whose distance to the boundary of $\Omega$ is at least equal to $\alpha$.
It is known that the set $\Omega_{\alpha, d}$ is a Cantor set (closed, perfect and nowhere dense) of full measure for sufficiently small $\alpha$.
The measure of $\Omega \setminus \Omega_{\alpha, d} $ is $\mathcal O(\alpha )$, which tends to zero as $\alpha \downarrow 0$, see Refs. \cite{Poschel01}, and \cite{HB90}.
Finally, we define the subset $$A_{\alpha, d}=  \omega^{-1} ( \Omega_{\alpha, d}) \subset A.$$
It is true that $A_{\alpha, d}$ is a Cantor set of full measure. The measure of the complement subset $A \setminus A_{\alpha, d}$ is of order $\mathcal O(\alpha )$ as $\alpha \downarrow 0$, see \cite{uni-kam}. The intersection of $p^{-1}(E)$ with $\mathbb T^2 \times A$ is of the form  $\mathbb T^2 \times \Gamma_a $, with a some curve denoted by $\Gamma_a$ in $A$, passing through $\xi_a$.

Now for $\alpha$ small enough, we have the quasi-periodic stability of the Diophantine invariant
Lagrangian tori in $\mathbb T^2 \times A_{\alpha, d} $, as the following theorem.
It is combined form the different known versions of the classical KAM theorem, see Refs. \cite{Broer91}, \cite{Poschel82}, \cite{Bost86}, and \cite{Poschel01}.

% In particular the paper cite 33 proved the smooth dependence, in the sense of Whitney, of the KAM tori on the frequencies.

\begin{theo}  \label{kam}

Assume that $p$ is local non-degenerate as Definition \ref{Kol}. Let $d >0$ fixed and $\alpha >0$ be small enough. Assume that $0
< \lambda   \ll \alpha ^2 $. Then there exists a map $\Phi_\lambda: \mathbb T^2 \times A
\rightarrow \mathbb T^2 \times A$ with the following properties:
\begin{enumerate}
            \item $\Phi_\lambda$, depending analytically on $\lambda$, is a $C ^\infty-$
                diffeomorphism onto its image, close to the identity map in the $C ^\infty-$
                topology.
            \item For each $\xi \in A$, the invariant Lagrangian torus $\Lambda_\xi = \mathbb
                T^2 \times \{\xi \} $ is sent, by $\Phi_\lambda$, to $ \Phi_\lambda (\Lambda_\xi) $ which is a Lagrangian
                torus, (close to $\Lambda_\xi$), denoted by $\Lambda_{\xi_ \lambda} $, and of the form $\Lambda_{\xi_\lambda}= \mathbb T^2
                \times \{\xi_\lambda \} $, with some $\xi_\lambda$ in a certain open subset
                $A_\lambda \subset A$, induced by $\Phi_\lambda$.

                Moreover, if $\xi \in A_{\alpha, d}$, then $\Lambda_{\xi_ \lambda}$, with
                $\xi_\lambda$ in a certain subset $A_{\alpha, d, \lambda} \subset A_\lambda $,
                is still uniformly Diophantine $H_{p_\lambda}-$invariant torus, called local KAM torus. The restricted map $\Phi_\lambda |_{\Lambda_\xi} $ on each Diophantine Lagrangian torus $\Lambda_\xi$, with $\xi \in A_{\alpha, d}$,
                conjugates the Hamiltonian vector field
                $H_p|_{\Lambda_\xi}= H_p(x, \xi)$, given in \eqref{Hp}, to the Hamiltonian vector
                field $H_{p_\lambda}|_{\Lambda_{\xi_ \lambda}}$, i.e., $ \Phi_\lambda
                |_{\Lambda_\xi}
                * H_p = H_{p_\lambda}$.

                In particular, if $\xi \in \Gamma_a $, then the torus $\Lambda_{\xi_ \lambda}
                \subset p_\lambda ^{-1}(E) \cap \ \mathbb T^2 \times \Gamma_{a, \lambda}$, with a
                certain curve $\Gamma_{a, \lambda} \subset A_\lambda$. Moreover, when $\xi \in
                \Gamma_a \cap A_{\alpha, d}$, the Liouville measure of the complement of the
                union of the KAM tori $\Lambda_{\xi_ \lambda} $, in $p_\lambda ^{-1}(E)$, is
                of order $\mathcal{O}(\alpha)$.
\end{enumerate}
\end{theo}
Note that in the previous theorem, we can see $\xi_\lambda$ as a smooth function of $ \xi \in A$ and with the parameter $\lambda$.
From this theorem, we obtain a Cantor family of positive measure of KAM tori, that is $A_{\alpha, d, \lambda}$, on each of which the Hamiltonian flow of the perturbed system $p_\lambda$ is quasi-periodic of constant vector field.

\begin{rema} \label{uni}
We would to cite a very interesting paper \cite{uni-kam}. It attested that the unicity of KAM tori is valid on a subset of full measure of Diophantine Liouville tori.
By eliminating a measure zero set from $A_{\alpha, d}$, one defines a subset $A^*_{\alpha, d}\subseteq  A_{\alpha, d} $ as the set of density points of $A_{\alpha, d}$, as following.
Let $D^*_{\alpha, d} \subseteq D_{\alpha, d}$ be the subset of density points of $D_{\alpha, d}$, and then let $A^*_{\alpha, d} = \omega^{-1} ( D^*_{\alpha, d} \cap \Omega_{\alpha, d})$. The subset $A^*_{\alpha, d}$ has all similar properties than ones of $A_{\alpha, d}$, see \cite{uni-kam}.
The unicity here in the sense that the local conjugacy map $\Phi_\lambda$ given in Theorem \ref{kam}, after restriction to $\mathbb T^2 \times A^*_{\alpha, d}$ is unique up to a torus translation. This unicity ensures the unicity of corresponding KAM tori in the phase space, that we shall discus in Sec. \ref{dis}.
\end{rema}

\subsection{Spectral monodromy of $P_{\varepsilon, \lambda}$ }   \label{end mono}
In this section, we use again notation of Sec. \ref{sp-lattice Sec}.

Let $p_1$ be an analytic function in a tubular neighborhood of $T^*M$, real on the real domain,
with $p_1(x, \xi) =  \mathcal O( m(Re(x,\xi))$ in the case when $M=\mathbb R^2$, and $p_1(x, \xi)=
 \mathcal O( \langle \xi \rangle ^m) $ in the manifold case.

Let \beq \label{P-lambda}
P_{\varepsilon, \lambda}, \ 0 < \lambda \ll 1, \ h \ll \varepsilon = \mathcal{O}(h^\delta),\ \textrm{with} \ 0< \delta <1, \eeq
 be a classical operator that is small perturbation of the selfadjoint operator
$P_{\lambda}:=P_{\varepsilon =0, \lambda}$, and with the $h-$leading term of the form
\beq  \label{p_lambda} p_{\varepsilon, \lambda}=p_\lambda +i \varepsilon q+ \mathcal O (\varepsilon
^2), \ \textrm{with} \ p_\lambda = p+ \lambda p_1. \eeq
Here we assume that $p, q$ are symbols satisfying all assumptions of Sec. \ref{sp-lattice Sec} and moreover, $p$ is globally non-degenerate, as the following.

Let $X= F^{-1}(U)$, where the momentum map $F$ and the set $U$ are given in Sec. \ref{sp-lattice
Sec}. Then the map $F|_X: X\rightarrow U$ defines a smooth locally trivial bundle, whose fibres are Liouville invariant Lagrangian tori. From the angle-action Theorem \ref{A-A}, we can cover $X$ by an atlas of angle-action charts $\{( V^c, \kappa^c ) \}_{c \in U}$.

\begin{defi}
We say that $p$ is globally non-degenerate (on $X$) if for an arbitrary such atlas of $X$, $p$ is local non-degenerate on every local angle-action chart $( V^c, \kappa^c )$, see Definition \ref{Kol}.
\end{defi}

% where $\kappa^c=(x^c, \xi^c): V^c= F^{-1}(U^c)\rightarrow \mathbb T^2 \times A^c$.
% As in \ref{frequence}, we define the local frequency map for $p$, $\omega^c:  A^c \rightarrow \mathbb R^2$, $\omega^c(\xi)= \frac{\partial (p\circ \kappa^{-1}) }{\partial \xi^c} (\xi)$.

% each local frequency map is a diffeomorphism onto its image, for every atlas of angle-action charts $\{V^c, \kappa^c \}_{c \in U}$ of $X$.

We will introduce a set, associated with a energy level $E$ of $p_\lambda$, which is similar to
the set of good values, given in Definition \ref{dn bonnes valeurs} of the integrable case.

Let $a=(E,G) $ be a point in $U$ such that $G$ is a good value in $ \mathcal{G}(\alpha,d, E)$, as
in Definition \ref{dn bonnes valeurs}. Then the corresponding torus $\Lambda_a  \simeq \Lambda_{ \xi_a}$, with $\xi_a \in
\Gamma_a$, satisfies the $(\alpha,d)-$Diophantine condition \eqref{dn alpha-d dioph}.

We assume moreover $0 < \lambda   \ll \alpha^2  $. Then by Theorem \ref{kam}, there exists a smooth
Cantor family of KAM tori close to $\Lambda_{ \xi_a}$, $\Lambda_{\xi_ \lambda} = \mathbb T^2 \times \{\xi_\lambda \}$, with $\xi_\lambda  \in A_{\alpha, d, \lambda}$, on which the $H_{p_\lambda}-$flow is quasi periodic of a uniformly Diophantine constant frequency, denoted by
$\omega_{\lambda}(\xi_\lambda)$. Therefore, over these KAM tori $\Lambda_{\xi_ \lambda}$, $p_\lambda$ become a function of only $\xi_\lambda$:
$$p_\lambda= p_\lambda (\xi_\lambda), \xi_\lambda  \in A_{\alpha, d, \lambda}. $$

% We assume that $p_\lambda$ extends to a smooth function of $\xi_\lambda$ on the whole domain  $A_\lambda$.

In particular, when $\xi \in \Gamma_a \cap A_{\alpha, d}$, then $\Lambda_{\xi_ \lambda} \subset
p_\lambda ^{-1}(E) \cap \ \mathbb T^2 \times  (\Gamma_{a, \lambda} \cap A_{\alpha, d, \lambda} )$, and with $\xi_ \lambda \in \Gamma_{a, \lambda} \cap A_{\alpha, d, \lambda} $.

Similarly as in \eqref{moyenne de q}, we define locally a smooth function $\langle q \rangle
_{\Lambda_{\xi_ \lambda} }$ of $\xi_\lambda  \in A_\lambda$, denoted by $\langle q \rangle (\xi_
\lambda )$, obtained by averaging $q$ over the tori $\Lambda_{\xi_ \lambda} $.
Then we have, in $C^1-$sense in $\xi_\lambda  \in A_\lambda$,
 $$\langle q \rangle (\xi_ \lambda ) = \langle q \rangle _{\Lambda_{\xi_ \lambda} }   \rightarrow
 \langle q \rangle _{\Lambda_{\xi } }=  \langle q \rangle (\xi), \ \textrm{as} \ \lambda \rightarrow 0 .$$
%  \langle q \rangle _{\Lambda_{\xi_ \lambda |_\lambda=0 } }=
% that is close to $ \langle q \rangle (\xi_a)= \langle q \rangle _{\Lambda_{\xi_a } }  $. \\
Hence, noticing the properties of the good value $G$, for every $\xi_\lambda \in \Gamma_{a, \lambda}$ and $\lambda$ small enough, we get
\beq
\label{dk1} \big | d \langle q \rangle (\xi_ \lambda ) \big | \geq \frac{\alpha}{2}. \eeq Notice that
we have also the differentials of $p_\lambda(\xi_ \lambda )$ and $\langle q \rangle (\xi_ \lambda
)$ in every $\xi_\lambda \in \Gamma_{a, \lambda} \cap A_{\alpha, d, \lambda} $ are $\mathbb R
-$linearly independent: \beq \label{dk2} \omega_{\lambda}(\xi_\lambda) \wedge d \langle q \rangle
(\xi_ \lambda ) \neq 0. \eeq
Let us define the set of (new) \textit{good values} for $P_{\varepsilon, \lambda}$,
\beq \label{K}
\mathcal{G_\lambda}(\alpha, E, G)= \bigcup_{\xi_\lambda \in \ \Gamma_{a, \lambda} \cap A_{\alpha, d,
\lambda}} \langle q \rangle (\xi_ \lambda ) = \bigcup \{\langle q \rangle (\xi_ \lambda ): \ \xi
\in \Gamma_a \cap A_{\alpha, d} \} .\eeq It is true that the measure of the complement of
$\mathcal{G_\lambda}(\alpha, E, G)$ in $\bigcup_{\xi_\lambda \in \Gamma_{a, \lambda}} \langle q
\rangle (\xi_ \lambda ) $ is small, is of order $\mathcal{O}(\alpha)$, when $\alpha$ is small and
$d$ is kept fixed.

If $K \in \mathcal{G_\lambda}(\alpha, E, G) $, then there exists a unique $\xi \in \Gamma_a \cap A_{\alpha,
d} $ such that $\langle q \rangle (\xi_ \lambda ) = \langle q \rangle _{\Lambda_{\xi_ \lambda} }= K$, and the corresponding KAM torus $\Lambda_{\xi_ \lambda} \subset p_\lambda ^{-1}(E)$ is still uniformly Diophantine $H_{p_\lambda}-$invariant Lagrangian torus. Moreover, the $H_{p_\lambda}-$flow on $\Lambda_{\xi_
\lambda} $ is quasi-periodic of the Diophantine constant frequency $\omega_{\lambda}(\xi_\lambda)$,
satisfying \eqref{dk1} and \eqref{dk2}. Therefore, these basis assumptions on the dynamic of
$H_{p_\lambda}$ allow us to carry out, microlocally near $\Lambda_{\xi_ \lambda} $, a construction
of the Birkhoff normal form for $P_{\varepsilon, \lambda}$.

\begin{rema}
The principe of the Birkhoff normalization is to use excessively canonical transformations near a Diophantine invariant torus (like $\Lambda_{\xi_ \lambda}$), to conjugate the operator $P_{\varepsilon, \lambda}$ to a new operator, whose total symbol is independent of angle variables $x$, and homogeneous polynomial to high order in $ (\xi, h, \varepsilon)$. This conjugation, with the help of the Egorov Theorem \cite{Egorov69}, is by means of analytic Fourier integral operators \cite{Duis11}.
The Diophantine condition of the torus ($\Lambda_{\xi_ \lambda} $) is indispensable for this construction. We refer to \cite{Hitrik07}, \cite{Charles08}, and \cite{Vu-Ngoc09} for the Birkhoff normal form procedure.
\end{rema}

So as a result of Ref. \cite{Hitrik07} (Section 7.3), one obtains asymptotic spectral results for the operator
$P_{\varepsilon, \lambda}$, that are similar to those of the operator $P_{\varepsilon}$ in the
integrable case.

\begin{theo}[\cite{Hitrik07}]  \label{v}
Let $P_{\varepsilon, \lambda}$ be the operator in \eqref{P-lambda}, with leading term
$p_{\varepsilon, \lambda}$ \eqref{p_lambda}, satisfying all assumptions given in this section. We
work in the regime $h \ll \varepsilon = \mathcal{O}(h^\delta)$, for $0< \delta <1 $, and $ \lambda > 0 $ small enough. Let $a=(E,G)$ be any point in $U$ such that $G \in \mathcal{G}(\alpha,d, E)$, with $\alpha > 0$ small enough, and $d>0$ fixed, as in Definition \ref{dn bonnes valeurs}. We
assume that $ \lambda   \ll \alpha^2$, and define the set $\mathcal{G_\lambda}(\alpha, E, G)$ as in
\eqref{K}.

For each $K \in \mathcal{G_\lambda}(\alpha, E, G) $, there exists a KAM torus $\Lambda_{\xi_
\lambda}=  p_\lambda ^{-1}(E) \cap \langle q \rangle ^{-1}(K) $ of $H_{p_\lambda}-$flow's frequency
$\omega _{\lambda}(\xi_\lambda)$ as already discussed above, and a canonical transformation
\begin{equation} \label{kappa vc}
\kappa_{\infty}=(x, \xi): \textrm{neigh} \ (\Lambda_{\xi_\lambda}, T^*M)\rightarrow
\textrm{neigh} \ (\xi=\xi_\lambda, T^*\mathbb T^2) ,
\end{equation}
mapping $\Lambda_{\xi_\lambda}$ to $ \mathbb T^2 \times \{\xi_\lambda \} $, such that
 \begin{equation} \label{lea. sym}
  p_\lambda \circ \kappa_{\infty}^{-1}= p_\lambda (x, \xi)=  p_{\lambda, \infty }(\xi) + \mathcal{O}(\xi-\xi_\lambda )^\infty,
 \end{equation}
where $p_{\lambda, \infty }$ is a smooth function, depends on $\xi$ only, and admits the Taylor expansion at $\xi_\lambda$ of the form
\begin{equation} \label{Taylor}
p_{\lambda, \infty }(\xi)= E+ \omega _{\lambda}(\xi_\lambda)\cdot (\xi-\xi_\lambda ) + \mathcal{O}(\xi-\xi_\lambda )^2.
\end{equation}
Let $\eta \in \mathbb Z^2$ be the Maslov index (see Def. \ref{ind}) and $S \in \mathbb R^2$ be the integral action, defined as in \eqref{act} of
the fundamental cycles of $\Lambda_{\xi_ \lambda} $, suitable with the canonical transformation \eqref{kappa vc}.
Then all the eigenvalues $\mu$ of
$P_{\varepsilon, \lambda}$ in the rectangle $R^{(E, K)}(\varepsilon,h)$ defined by \eqref{cua so} with $K$ instead of $G$, are given as the image of a portion of $h \mathbb Z^2$, with modulo $\mathcal O(h^\infty)$, by a smooth function $P_\lambda (\xi, \varepsilon;h)$ of $\xi$ in a neighborhood of $(\xi_\lambda, \mathbb R^2)$ and $\varepsilon, h$ in neighborhoods of $(0, \mathbb R)$:
 \begin{equation} \label{eigenvalues-with lambda}
   \sigma(P_{\varepsilon, \lambda}) \cap R^{(E, K)}(\varepsilon,h) \ni \mu =
   P_\lambda \Big( \xi_\lambda+ h(k-\frac{\eta}{4})-\frac{S}{2 \pi};\varepsilon, h\Big) + \mathcal O(h^\infty), \  k \in \mathbb Z^2.
  \end{equation}
Moreover, $P_\lambda $ is real valued for $\varepsilon =0 $, admits an asymptotic expansion in $(\xi, \varepsilon, h)$ of the form \eqref{symbole normal}. In particulary, the $h-$leading term of $P_\lambda$ is of the form
\beq \label{prin normal-lambda} p_{0, \lambda}(\xi, \varepsilon) =
p_{\lambda, \infty }(\xi)+ i \varepsilon \langle q \rangle (\xi) + \mathcal O(\varepsilon ^2), \eeq
where $p_{\lambda, \infty }(\xi)$ is given in \eqref{lea. sym}, and $ \langle q \rangle (\xi)$ is the expression of the averaging of $q$ over the tori $\Lambda_\xi$ close to $\Lambda_{\xi_\lambda} $, in the previous coordinates.
\end{theo}

We notice that the canonical transformation $\kappa_{\infty} $ in \eqref{kappa vc}, which satisfies the properties \eqref{lea. sym} and \eqref{Taylor}, is locally valid for every KAM torus in a neighborhood of $\Lambda_{\xi_\lambda}$. Hence, the construction in the above theorem is locally valid for any good value given $a=(E, G) \in U$, and for $K \in \mathcal{G_\lambda}(\alpha, E, G)$.

We can see that the results in \eqref{eigenvalues-with lambda} and \eqref{prin normal-lambda} are very similar to those in \eqref{eigenvalues-E} and \eqref{prin normal-E} of Sec. \ref{sp-lattice Sec}. Hence the way to construct a spectral asymptotic pseudo-lattice here is the same as of Sec. \ref{sp-lattice Sec}. Therefore, in order to avoid repetitions, we shall skip some technical details.

Eq. \eqref{eigenvalues-with lambda}, with the aid of \eqref{prin normal-lambda}, provides a micro-chart, denoted by $f_\lambda$, of the spectrum of $P_{\varepsilon, \lambda}$ on the rectangle $R^{(E, K)}(\varepsilon,h)$. This micro-chart satisfies \eqref{new hk} and \eqref{f tilde}, with $P$ replaced by $P_\lambda$ and $G$ by $K$. It is true that there exists locally a Cantor family of such micro-charts. Moreover, noticing the fact that the union of the invariant KAM tori is of full measure, thus this family is a pseudo-chart.

In the reduced form, the asymptotic expansion in $(\varepsilon,\frac{h}{\varepsilon})$ of $\widetilde{f_\lambda}:= f_\lambda \circ \chi$, with $\chi$ given by \eqref{chi}, for the $C ^\infty-$ topology in a neighborhood of $(E,K)
\in \mathbb R^2$, satisfies $$\widetilde{f}_{\lambda}:= \widetilde{f}_{\lambda 0} + \mathcal
O(\varepsilon, \frac{h}{\varepsilon}), $$ uniformly for $h, \varepsilon$ small and $h \ll
\varepsilon $. Here $\widetilde{f}_{\lambda,0}$ is the leading term of $\widetilde{f}_{\lambda}$.
Moreover, we have also a similar result as one in \eqref{pre terme f tilde} that
\begin{equation}   \label{z}
  \widetilde{f_\lambda}_0= \tau_a+ \psi^{-1},
\end{equation}
where \begin{equation} \label{psi}
 \psi(\xi) = ( p_{\lambda , \infty } (\xi) , \langle q \rangle (\xi)  ),
\end{equation}
with $\xi$ in a neighborhood of $(\xi_\lambda, \mathbb R^2)$, and $\tau_a$ is a locally constant in $a \in U$.

Thus we can state that the spectrum of $P_{\varepsilon, \lambda}$ on the domain $U(\varepsilon)$ satisfies all hypothesis of an asymptotic pseudo-lattice, according Definition \ref{pseu-lattice}.
In conclusion, we have:
\begin{theo}
$(\sigma( P_{\varepsilon, \lambda}), U(\varepsilon) )$ is an asymptotic pseudo-lattice.
\end{theo}
Therefore, we can define the spectral monodromy of the operator $P_{\varepsilon, \lambda}$ due to the monodromy of the asymptotic pseudo-lattice $(\sigma( P_{\varepsilon, \lambda}), U(\varepsilon) )$, as discussed in Sec. \ref{sp-lattice Sec} (see also \cite{QS14} for more detail). That is the class $[\mathcal M_{sp}] \in \check{H}^1(U(\varepsilon), GL(2, \mathbb Z) )$
in the \v{C}ech cohomology group, defined from the $1$-cocycle of integer linear transition maps
$\{M_{ij} \}_{i, j \in \mathcal{J} } $, given in \eqref{transition-pseu}. In the context of this section, we have
\begin{equation}  \label{d}
M_{ij}= d \big ( \widetilde{f}_{i \lambda,  0} \circ (\widetilde{f}_{j \lambda,  0})^{-1} \big ), \ i, j \in \mathcal{J}.
\end{equation}

Note that these $M_{ij}$ don't depend on $\lambda$ small enough, by the discreteness of the integer
group $GL(2, \mathbb Z) $. So the spectral monodromy of $P_{\varepsilon, \lambda}$ is independent of any small perturbation
caused by $\lambda$. It is known that it is already independent of classical parameters $h, \varepsilon $.

%%%%%%%%%%%%%%%%%%%%%%%%%%%%%%%%%%%%%%%%%%%%%%%%%%%%%%%%%%%%%%%%%%%%%%%%%%%%%%%%%%%%%%
%%%%%%%%%%%%%%%%%%%%%%%%%%%%%%%%%%%%%%%%%%%%%%%%%%%%%%%%%%%%%%%%%%%%%%%%%%%%%%%%%%%%%%

\subsection{Discussions}  \label{dis}
% Conclusion

The spectral monodromy is well defined for small non-selfadjoint perturbations of a selfadjoint classical operators in dimension $2$, admitting a quasi-integrable unperturbed principal symbol, as discussed in Sec. \ref{end mono}. On the other hand, there exists a interesting result concerning a geometric invariant of quasi-integrable systems, given by Broer and co-worker, see Ref. \cite{Broer07}.
It showed the construction of a global Whitney smooth conjugacy (global KAM) between a quasi-integrable system and an integrable one. That means the global quasi-periodic stability of invariant Lagrangian tori in phase space, which allows them to define a monodromy for KAM tori. For simplicity, we say it as the KAM monodromy. Naturally, we discuss in this section the relationship between the spectral monodromy and the KAM monodromy.

We consider a $H_p-$invariant Lagrangian torus $\Lambda_a= F^{-1}(a) \subseteq X $, with $a \in U$. Let $(V,\kappa) $ be a local angle-action chart near $\Lambda_a$, and $\Phi_\lambda$ be the corresponding local KAM conjugacy given by Theorem \ref{kam}.
We define on the phase space the Lagrangian torus $$\Lambda_{a, \lambda}= \kappa^{-1}\circ \Phi_\lambda \circ \kappa (\Lambda_a) \subseteq X .$$
A interesting result from \cite{Broer07} showed that if the action coordinates $\xi_a \in A^*_{\alpha, d}$, where the set $A^*_{\alpha, d}$ is defined in Remark \ref{uni}, then the corresponding KAM torus $\Lambda_{a, \lambda}$ in the phase space doesn't depend on the choice of such angle-action chart $(V,\kappa) $.
In this way, we can obtain all invariant KAM tori on $X$. In fact, there exists a nowhere dense subset $C \subset U $ and the measure of the set $U \setminus C$ tends to $0$ as $\lambda$ tends to zero. Accordingly, the set of Liouville tori $F^{-1}(C) \subset X$ is also nowhere dense of full measure.
Moreover, by gluing together local KAM conjugacies with the help of a partition of unity, one obtains a global conjugacy as a finite convex linear combination of local conjugacies. There exists so a $C^\infty-$diffeomorphism (global conjugacy) $\Phi$ on the entire space $X$, whose restriction on $F^{-1}(C)$ conjugates each $\Lambda_a$, with $a \in C$, to the $H_{p_\lambda}-$invariant KAM torus $\Lambda_{a, \lambda}$. This conjugacy provides an equivalence (smooth in the sense of Whitney) between the integrable system and its small perturbation on a nowhere dense set of large measure.

Hence the union of all \textit{global KAM tori} in $X$ is the nowhere dense subset $\bigcup_{a \in C} \Lambda_{a, \lambda} \subseteq X$. Moreover, the measure of KAM tori in $X$ is large,
$$\mu(X \setminus \bigcup_{a \in C} \Lambda_{a, \lambda})  = \mathcal O( \alpha ) \mu (X). $$
Thus one obtains a Whitney smooth foliation of invariant KAM tori over a nowhere dense set with complement of small measure.

We notice here that the functions $p_\lambda$ and the average function $\langle q \rangle = \langle q \rangle _{\Lambda_{a, \lambda} }$ are Whitney smooth, independent (in seeing \eqref{dk2}), and in involution when restricted to the nowhere dense set of large measure of the global KAM tori.
One says that this is the aspect of \textit{integrability for perturbed systems}.
In the context $n=2$, this result which is global, is stronger than one of Ref. \cite{Poschel82}, which proved locally the existence of a system with those properties. On the other hand, the previous system found in our work is different from one of the last paper, which is locally given by the components of the frequency $\omega _{\lambda}(\xi_\lambda)$ of $H_{p_\lambda}-$flow on local KAM tori, expressed as functions on the phase space.

It is know that on the Liouville tori of integrable systems, the classical monodromy is well defined by Duistermaat
\cite{Duis80}. Whether one can define such a geometric invariant for KAM tori of quasi-integrable perturbed systems? The answer is positive if  unperturbed integrable systems are globally non-degenerate, given in \cite{Broer07}.
The KAM monodromy is defined as the obstruction against global triviality of a $\mathbb Z^2-$bundle over the whole base $U$, denoted by $\mathscr F$, which is an extension from the $\mathbb Z^2-$bundle $H_1(\Lambda_{a, \lambda}, \mathbb Z) \rightarrow a \in C$.
Here $H_1(\Lambda_{a, \lambda}, \mathbb Z)$ is the first homology group of $\Lambda_{a, \lambda}$.

Let $\xi_\lambda$ be local action coordinates, given by \eqref{kappa vc}, of the global KAM tori $\Lambda_{\lambda} \simeq \Lambda_{\xi_\lambda}$. We have $\psi(\xi_\lambda)= (p_{\lambda}( \xi_\lambda), \langle q \rangle ( \xi_\lambda ) )$, where the map $\psi$ is defined by \eqref{psi}. So the map $\psi^{-1}$ give the same local action variables of $\Lambda_{\lambda}$. It is classical that there is a one-to-one correspondence between the compositions of $d ( \psi ^{-1})$ and periodic flows of period $1$ on $\Lambda_{\lambda}$, which form a basis of $H_1(\Lambda_{\lambda}, \mathbb Z)$, see \cite{Duis80}.

On the other hand, we look at now the transition maps from the spectral monodromy $[\mathcal M_{sp}]$ of $P_{\varepsilon, \lambda}$. Using \eqref{z} and \eqref{d} we get
$$ d \big ( \psi_i^{-1} \circ \psi_j \big ) = M_{ij}  \in  GL(2, \mathbb Z) .$$

This result asserts that the above homology bundle $\mathscr F $ has an integer linear structure group. The transition maps between local trivializations of the bundle are $ \{ {}^tM_{ij}^{-1} \} $.
Therefore, the spectral monodromy recovers completely the KAM monodromy of quasi-integrable systems. Moreover, we have a similar result as one of the integrable case.
\begin{theo}
The spectral monodromy of $P_{\varepsilon, \lambda}$ is the adjoint of the KAM monodromy.
\end{theo}

In conclusion for this paper, the spectral monodromy is well defined directly from the spectrum of small non-selfadjoint perturbations of a selfadjoint operators admitting, either a completely integrable (semi-)classical principal symbol, or a quasi-integrable one with a
globally non-degenerate condition. In both cases, the spectral monodromy allows to recover, either
the classical monodromy of the Liouville invariant tori, defined by Duistermaat \cite{Duis80}, or the
monodromy of the KAM invariant tori, defined by Broer and co-workers \cite{Broer07}. That illustrates
clearly the principle idea of semiclassical analysis that classical mechanics is the
semiclassical limit of quantum mechanics.

% truong hop dac biet khi p,q commute thi ....

% Truong hop dac biet khi perturbation is still integrable, then the same classical

% voi d co dinh, tinh chat Uniformly Dio la ko phu thuoc vao local coord, xem Broer...

% see also 37 for the classical Birkhoff construction in the non-resonant case, near a stable equilibrium point.

% the Bir normal form construction: For the Bir, we refer to

% chu y rang de co no thoa man cac gia thiet ve elippticity o vo cung....cac gai thiet chung doi voi toan tu moi cung van dung tuong tu toan tu cu.

% AP de dn monodromy for classical resonances.

% Acknowledgment: tham khao trang 10 cua Hitrik 07.

% Duistermaat da xd monodoromie clas cho he integralbe,   Broer da xd monodromie cho tap hop cac Dio tori KAM bang cach....

%%%%%%%%%%%%%%%%%%%%%%%%%%%%%%%%%%%%%%%%%%%%%%%%%%%%%%%%%%%%%%%%%%%%%%%%%%%%%%%%%%%%%%%%%%%%%%%%%%%%%%%%%%%%%%%%%%%%%%%%%%%%%%%%%%%%%%%%%%%%%%%%%%%%%%%%%%%%%%%%%%%%%%%%%%%%%%%%%%%
%%%%%%%%%%%%%%%%%%%%%%%%%%%%%%%%%%%%%%%%%%%%%%%%%%%%%%%%%%%%%%%%%%%%%%%%%%%%%%%%%%%%%%%%%%%%%%%%%%%%%%%%%%%%%%%%%%%%%%%%%%%%%%%%%%%%%%%%%%%%%%%%%%%%%%%%%%%%%%%%%%%%%%%%%%%%%%%%%%%
%%%%%%%%%%%%%%%%%%%%%%%%%%%%%%%%%%%%%%%%%%%%%%%%%%%%%%%%%%%%%%%%%%%%%%%%%%%%%%%%%%%%%%%%%%%%%%%%%%%%%%%%%%%%%%%%%%%%%%%%%%%%%%%%%%%%%%%%%%%%%%%%%%%%%%%%%%%%%%%%%%%%%%%%%%%%%%%%%%%

\bigskip

\end{document}